# From localized dryout to convective elongated vapor structures: Reynolds number effects on boiling transition in a rectangular mini-channel

Qi Wang[1, 2], Xin Wang[1], Mingze Wang[1, 2], Yifei Guan[3], Kang Luo[1, 2], Jian Wu[1*], Wei Wang[1, 2], Alberto T. Pérez[4]

[1.] *School of Energy Science and Engineering, Harbin Institute of Technology, Harbin 150001, China*

[2.] *Suzhou Research Institute of HIT, Suzhou 215104, China*

[3.] *Department of Mechanical Engineering, Union College, Schenectady, NY, USA*

[4.] *Department of Applied Physics III, ETSI, Universidad de Sevilla, Sevilla, Spain*

## Abstract

Three-dimensional conjugate simulations were conducted to investigate saturated flow boiling in a rectangular mini-channel, with particular emphasis on the role of inlet Reynolds number on boiling mode selection and transition. A C++ based open-source numerical framework was employed, incorporating a physically informed multi-site nucleation model by coupling a nucleation site density correlation with a Halton-sequence based spatial allocation strategy. Two distinct *Re*-dependent transition pathways were identified. At low *Re*, boiling transition is mainly associated with localized dryout development associated with upstream active boiling and progressive downstream liquid starvation. At high *Re*, the transition is characterized by convective stretching and reorganization of vapor structures, through which elongated vapor slugs evolve into localized vapor films and eventually approach full surface vapor coverage. The global heat transfer characteristics and peak heat transfer capacity are further interpreted in conjunction with boiling mode transition, clarifying the respective roles of wall dryout and volumetric vapor fraction in heat transfer deterioration. Among all cases, *Re*=2000 provides the most favorable overall thermal response. Overall, within the rectangular mini-channel configuration and operating range considered in this study, *Re* is closely associated with vapor organization, boiling transition, wall dryout, and global heat transfer performance.

## I. INTRODUCTION

With the rapid rise of heat flux and temperature nonuniformity in advanced electronic and compact energy systems, thermal management approaches are increasingly required to deliver both high heat dissipation capability and local thermal reliability. In this context, flow boiling in micro-/mini-channels remains one of the most promising cooling strategies because it combines large latent heat transport capacity with compact integration potential. Recent reviews have emphasized continued advances in both the physical understanding and numerical modeling of microchannel flow boiling, while newly reported studies have further demonstrated its relevance to ultra high heat flux chip cooling and other compact thermal systems[1-3].

* Corresponding author. Tel.: +86 18321925448; E-mail address: jian.wu@hit.edu.cn.

Among the governing parameters, channel size, imposed heat flux, and inlet mass flux have long been recognized as key factors controlling vapor confinement, flow pattern evolution, and boiling heat transfer in microchannels. Early experimental studies by Qu and Mudawar[4] demonstrated the distinctive characteristics of two phase flow boiling in microchannel heat sinks and highlighted the strong coupling among pressure drop, vapor generation, and heat transfer. Harirchian and Garimella[5, 6] subsequently showed that channel dimension and mass flux strongly affect local boiling behavior and proposed comprehensive regime maps and transition criteria for microchannel flow boiling. Thome et al.[7] further emphasized the multiscale nature of confined two phase flow patterns and the need for mechanistic flow pattern based interpretation. Collectively, these studies have made it clear that boiling heat transfer in confined rectangular channels cannot be understood solely from averaged thermal quantities, because the underlying vapor structures and their transitions play a decisive role in the observed thermal response.

Recent experimental studies have continued to enrich this picture, especially for rectangular mini-/micro-channel configurations that are directly relevant to practical heat sinks. For example, Wang et al.[8] and Fang et al.[9] investigated saturated flow boiling of R123 in mini-channel heat sinks and showed that flow pattern development remains tightly linked to heat transfer performance. Cheng and Wu[10] further demonstrated that geometry-dependent confinement substantially modifies flow boiling characteristics in rectangular microchannels. These recent studies confirm that even when the working fluid and geometry are changed, the interplay between confinement, vapor organization, and inlet flow condition remains central. However, the majority of experimental studies are still primarily oriented toward boiling curves, pressure drop, and regime maps, while a mechanism resolved understanding of how the inlet Reynolds (*Re*) number reorganizes vapor structures and changes the transition pathway between dominant boiling modes (which refer to the characteristic vapor organization and boiling-state features that control the heat transfer behavior during the boiling transition) remains comparatively limited, particularly when local wall thermal response is considered together with global heat transfer metrics.

In parallel with experiments, numerical simulations have become increasingly important for resolving interfacial evolution (which refers to the transient development of the liquid–vapor interface, including bubble nucleation, growth, deformation, coalescence, streamwise elongation, and vapor film formation), wall heat transfer (which refers to the local and time-averaged thermal response at the heated conjugate boiling surface, including wall heat flux, Nusselt number, wall wetting/dryout condition, and heat transfer deterioration), and dryout dynamics that are difficult to measure directly. Recent high fidelity studies have examined bubble growth under conjugate heat transfer[11], three-dimensional (3D) subcooled flow boiling in rectangular mini-channels[12], conjugate flow boiling in non-circular microchannels[13], and critical heat flux triggering in vertical rectangular mini-channels[14]. Other studies have addressed the effects of surface roughness[15], reviewed numerical approaches across scales[1], and explored more specialized geometries and operating conditions such as micro-pin-fin structures, multiple ultra high heat flux sources, expanding microchannels, stratified boiling, and channel orientation effects[16-18]. These works have significantly advanced understanding of interfacial dynamics, thin film evaporation, and local wall heat transfer. Nevertheless, much of the existing numerical literature focuses on subcooled boiling, a single seeded bubble, a small number of prescribed nucleation sites, modified channel geometries, or CHF oriented problems, rather than on transition resolved saturated flow boiling in a straight rectangular mini-channel where the inlet *Re* is treated as the primary mode-selecting parameter.

A second issue concerns the physical representation of nucleation in interface-resolved boiling simulations. Although continuum scale phase change formulations, such as the model of Hardt and Wondra[19], provide a consistent framework for interfacial mass and energy transfer, the treatment of heterogeneous nucleation remains a major source of modeling uncertainty. Recent studies have shown that the choice of nucleation model can materially affect the predicted bubble behavior and flow boiling heat transfer[20]. At the same time, many simulations still rely on a few manually prescribed embryos or predetermined nucleation sites for numerical convenience, which is acceptable for studying isolated bubble behavior but less suitable for investigating boiling regime transition over a wide range of operating states. Low-discrepancy spatial allocation strategies have been adopted in high fidelity pool boiling simulations to improve site distribution and repeatability[21, 22], but their systematic integration with nucleation site density based activation in 3D conjugate simulations of saturated rectangular mini-channel flow boiling remains scarce.

In interface-resolved simulations of flow boiling, the treatment of wall nucleation is important because the onset of boiling and the number of active vapor embryos cannot be determined solely by the interfacial phase-change model. If the nucleation process is represented only by a few manually prescribed embryos or fixed nucleation sites, the simulation may capture isolated bubble growth but may become sensitive to the assumed number and location of nucleation sites when multi-site boiling transition is considered. To reduce such arbitrariness, active nucleation site density models have been developed to relate the number of active sites to wall and surface conditions. For example, Hibiki and Ishii[23] proposed a widely used active nucleation site density model in which the active site density is related to the critical cavity size and contact angle, and validated it using data from both pool boiling and convective flow boiling systems. More recently, Li et al.[24] developed a nucleation site density correlation based on a parametric analysis of existing experimental data, in which wall superheat, pressure, and contact angle are explicitly considered, and further demonstrated its preliminary applicability in CFD simulations of boiling flows. Chen et al. also showed that the heterogeneous nucleation treatment can significantly influence the predicted vapor distribution and heat transfer behavior in mini-channel flow boiling simulations. These studies indicate that nucleation modeling is not merely a numerical triggering treatment, but an important factor affecting the predicted boiling transition and heat transfer characteristics. Therefore, in the present work, the active nucleation site density model of Li et al.[24] is adopted to determine the number of active nucleation sites, and a Halton low-discrepancy sequence is further used to distribute these sites over the heated wall in a deterministic and reproducible manner.

For the present problem, this gap is particularly important. Inlet Reynolds number does not merely change the magnitude of convection; it also changes bubble residence, vapor stretching, coalescence behavior, liquid replenishment, and therefore the competition between dryout expansion and rewetting recovery. As a result, different *Re* may lead not only to different average heat transfer levels, but also to qualitatively different boiling modes and transition pathways. From the viewpoint of thermal design, such differences are crucial because a configuration with seemingly favorable average performance may still suffer from more severe local dryout and higher wall hot spots. Therefore, a framework that can simultaneously capture multi-site nucleation, evolving vapor organization, and local wall thermal response is needed to examine how inlet *Re* affects boiling-transition behavior in saturated rectangular mini-channel flow boiling.

Motivated by the above considerations, the present study performs 3D conjugate simulations of saturated flow boiling in a rectangular mini-channel, with particular emphasis on how inlet *Re* affects the dominant boiling mode and its associated transition process. A physically informed multi-site nucleation framework is employed by coupling a nucleation site density model with a Halton based spatial allocation strategy. This treatment allows reproducible activation of multiple nucleation sites while avoiding the over simplification associated with a few manually imposed nuclei and reducing the clustering sensitivity of purely random placement. On this basis, the study aims to identify *Re*-dependent boiling modes, clarify their transition pathways along the boiling curve, and connect these transitions to global heat transfer characteristic. In particular, two dominant mode families are revealed: a localized dryout dominated mode under lower *Re* conditions and a convection dominated elongated vapor mode under higher *Re* conditions. The central objective of this work is thus not simply to compare average boiling performance at different inlet *Re*, but to examine how inlet inertia is associated with vapor-structure reorganization, dryout competition, and differentiated global heat transfer responses in the present simulations.

Furthermore, the present configuration and the developed solver have been carefully designed to provide a reliable baseline and a physically consistent numerical framework for subsequent studies on electrically active control of boiling heat transfer. This framework builds on our previous work in electric field enhanced heat transfer[25-27] and supports further studies on active electric field control of boiling heat transfer. In particular, by establishing a transition resolved understanding of flow boiling in the absence of external fields, the current work lays the foundation for future investigations on electric field assisted boiling, where the interaction between interfacial dynamics and electrohydrodynamic effects can be systematically explored. The remainder of this work is organized in the following way. The physical model, the governing equations, and the numerical method are described in Sec. II., the validation of the solver is shown in Sec. III. Section IV presents a comprehensive discussion of the results. Section V contains the concluding remarks.

## II. MATHEMATICAL MODEL AND NUMERICAL METHOD

### A. Physical model

In the present study, a conventional rectangular mini-channel with a commonly used geometry was adopted. Its configuration and dimensions are illustrated in Fig.1, the computational domain consists of two parts, namely a fluid region and a solid region. The fluid region has a square cross section of 3 mm×3 mm and a length of 60 mm, while the copper substrate beneath the channel has a thickness of 1 mm. It should be noted that changes in channel width, height, aspect ratio, heated length, or substrate thickness may influence vapor confinement, liquid replenishment, conjugate heat spreading, and dryout evolution. Therefore, the boiling-transition pathways discussed in this work should be interpreted within the present geometry, working fluid, boundary conditions, and operating range, rather than as a universal criterion for all mini-/micro-channel configurations.

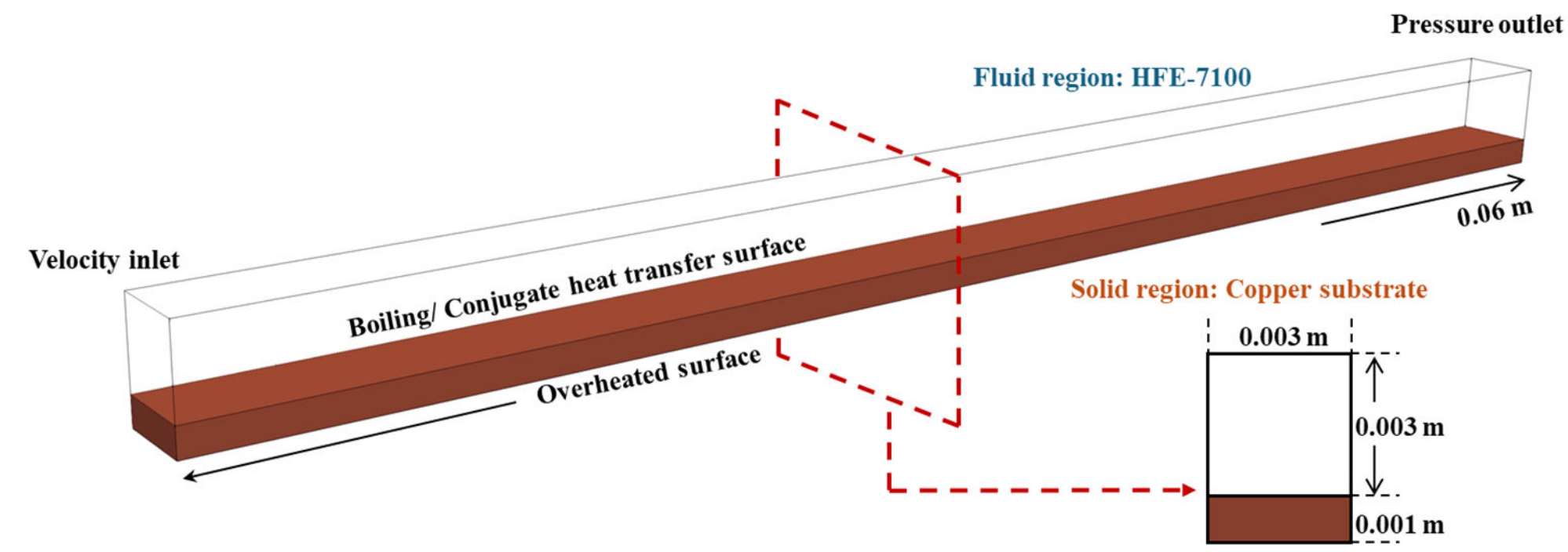

Fig.1 Graphical representation of the physical configuration of the rectangular mini-channel

The working fluid in this work is HFE-7100, a dielectric fluorinated liquid widely used in electronics cooling applications. The thermophysical properties of its liquid and vapor phases are listed in Table.1. These property values were determined based on a combination of data reported in the studies of Zihuan Ma et al.[28] and the 3M product data sheet. The solid substrate was made of copper and its thermophysical properties are also summarized in Table.1.

Table.1 Material properties of HFE7100 liquid/vapor and copper

| Physical properties | HFE7100 liquid | HFE7100 vapor | copper |
|---|---|---|---|
| Density, kg/m$^3$ | 1420 | 11.5 | 8000 |
| Viscosity, Pa·s | $4 \times 10^{-4}$ | $1.32 \times 10^{-5}$ | / |
| Specific heat capacity, J/(kg·K) | 1263 | 870 | 450 |
| Thermal conductivity, W/(m·K) | 0.058 | 0.01 | 80 |

| Saturation temperature, K | 339 | / | / |
|---|---|---|---|
| Latent heat, J/kg | 111600 | / | / |
| Surface tension, N/m | 0.00943 | | |
| Molar mass, kg/mol | 0.25 | 0.25 | 50 |

For the fluid region, the left boundary was specified as a constant velocity inlet, while the right boundary was defined as a constant pressure outlet with the pressure fixed at 1 atm. The interface between the fluid and the copper substrate served as the conjugate heat transfer interface, where boiling phase change occurred and nucleation sites were prescribed. All the remaining walls were assumed to be no-slip and adiabatic. At the bottom surface of the copper substrate, a constant superheat temperature condition was imposed, so that heat was conducted upward through the solid wall and transferred to the fluid. Boiling phase change then took place at the conjugate heat transfer interface. The corresponding boundary conditions are summarized in Table.2.

Table.2 Boundary conditions

| | Velocity boundary | Pressure boundary | Temperature boundary |
|---|---|---|---|
| Velocity inlet | $u = u_0$ | $\partial P/\partial x = 0$ | $T = T_{sat}$ |
| Pressure outlet | $\partial P/\partial x = 0$ | $P = P_{atm}$ | $\partial T/\partial x = 0$ |
| Top wall (Liquid region) | noSlip | $\partial P/\partial y = 0$ | $\partial T/\partial y = 0$ |
| Coupled heat transfer surface | noSlip | $\partial P/\partial y = 0$ | / |
| Bottom wall (Solid region) | / | / | $T = T_0$ |
| Other | noSlip | $\partial P/\partial z = 0$ | $\partial T/\partial z = 0$ |

The numerical solver employed in this study incorporates the governing equations for fluid flow and heat transfer in both the fluid and solid regions, including the Navier-Stokes equations and energy equations. In addition, a phase change model proposed by Hardt and Wondra is adopted to describe the boiling process. A nucleation site density model developed by Li et al.[24], together with a spatial distribution strategy for nucleation sites, is also implemented to account for multi-site nucleation behavior.

## B. Navier-Stokes equations and energy equations

For the fluid region, the governing equations consist of the Navier-Stokes equations with phase change effects and the corresponding energy equation：

$$\rho \nabla \cdot \mathbf{u} = \dot{\rho} \quad (1)$$

$$\rho\left(\frac{\partial \mathbf{u}}{\partial t} + (\mathbf{u} \cdot \nabla)\mathbf{u}\right) = -\nabla P + \nabla \cdot \tau + \gamma \kappa \nabla \alpha + f_g \quad (2)$$

$$\frac{\partial}{\partial t}\left(\rho C_{pl} T\right) + \nabla \cdot \left(\rho C_{pl} \mathbf{u} T\right) = \nabla \cdot (k_l \nabla \mathrm{T}) + \dot{h} \quad (3)$$

Where $f_g$ denotes the body force induced by gravity, $\dot{h}$ accounts for the heat source associated with phase change. The viscous stress tensor is defined as:

$$\tau_{ij} = \mu\left(\frac{\partial u_i}{\partial x_j} + \frac{\partial u_j}{\partial x_i}\right) \quad (4)$$

the curvature of the interface $\kappa$, expressed as:

$$\kappa = -\nabla \cdot \left(\frac{\nabla \alpha}{|\nabla \alpha|}\right) \quad (5)$$

For the solid region, only the heat conduction equation is considered:

$$\frac{\partial}{\partial t}\left(\rho C_{ps} T\right) - \nabla \cdot (k_s \nabla \mathrm{T}) = 0 \quad (6)$$

### C. Phase change model

In the present study, the Volume of Fluid (VOF) method is employed for interface capturing. The VOF approach is adopted because it can directly resolve the liquid-vapor interface topology and near-wall dryout evolution, whereas two-fluid models require additional closure relations for these interfacial processes. The flow of two immiscible fluids is simulated by solving and tracking the volume fraction $\alpha(\mathbf{x},t)$ in each computational cell. Both phases are assumed to be incompressible and immiscible. The liquid volume fraction $\alpha$ is governed by the following conservative transport equation:

$$\frac{\partial \alpha}{\partial t} + \nabla \cdot (\alpha \mathbf{u}) = S_\alpha \quad (7)$$

In the present solver, the convective transport of $\alpha$ is discretized using the geometric isoAdvector method. The volume fraction $\alpha(\mathbf{x},t)$ is defined as the ratio of the liquid volume to the total volume of a computational cell, and the distribution of thermophysical properties $C$ within the domain is then determined based on the local volume fraction.

$$C = C_1 \alpha + C_2 (1 - \alpha) \quad (8)$$

The phase change process during boiling is modeled using the numerical approach proposed by Hardt and Wondra[19]. Following their work, phase change is introduced through a volumetric mass source term that is consistently coupled with the energy equation. The interfacial evaporation heat flux density is described by a linear kinetic relation:

$$j_e^h = c_e(T_i - T_{sat}) \tag{9}$$

$$c_e = \frac{2\chi_e}{2-\chi_e}\frac{h_e^2}{\sqrt{2\pi R}}\frac{\rho_v}{T_{sat}^{3/2}} \tag{10}$$

With the evaporation heat transfer coefficient $c_e$ given by the Schrage/Tanasawa type expression. $j_e^h$ is evaporation heat flux density, $\chi_e$ is the evaporation coefficient, which is set to 1.0 in this work. $h_e$ is the enthalpy of evaporation, R the gas constant for liquid. Eq. (10) describes the rate of evaporation or condensation at an already resolved liquid–vapor interface, whereas the nucleation site density model determines where and how many vapor embryos are introduced on the heated wall. The corresponding interfacial mass flux is $j_e = j_e^h/h_e$. A localized phase change rate field is constructed on the Eulerian grid as:

$$\varphi_0 = N j_e \alpha|\nabla\alpha| \tag{11}$$

Here, $\alpha$ is introduced to ensure that only field values on the liquid side of the interface determine the evaporation mass flux.The factor N is a normalization factor used when the interfacial mass flux is converted into a volumetric phase-change source term in the finite volume cells:

$$N\int_\Omega^1 \alpha|\nabla\alpha|d\Omega = \int_\Omega^1 |\nabla\alpha|d\Omega \tag{12}$$

To avoid numerical stiffness, $\varphi_0$is regularized by solving a Helmholtz type smoothing equation for an additional scalar field:

$$\nabla^2\varphi = \frac{1}{\Delta t D}(\varphi - \varphi_0) \tag{13}$$

where *D* controls the smoothing thickness. The resulting smoothed field u is used to define the mass source term

$$\dot{m} = N_v(1-\alpha)\,\varphi - N_l\alpha\,\varphi \tag{14}$$

where the normalization factors $N_v$ and $N_l$ are computed such that (i) the total evaporated mass equals the total generated vapor mass and (ii) the prescribed interfacial mass flux is recovered. More details of this model can be found in Hardt and Wondra[19].

**D. Nucleation density model and spatial implantation method**

The incorporation of a nucleation model is essential for reproducing the complete boiling curve in numerical simulations. In some existing studies, the nucleation process is not explicitly modeled, and simulations are therefore often limited to either a restricted nucleate boiling regime with a prescribed number of active sites or fully developed film boiling. As a result, the full transition from single phase conduction to nucleate boiling, transition boiling, and film boiling cannot be captured. One of the key features of the present study is the implementation of a physically based nucleation site density model, coupled with a rational spatial distribution strategy, which enables the numerical reproduction of the complete transition process from nucleate boiling to film boiling. In this study, the nucleation site density model proposed by Li et al.[24], obtained based on experimental fitting, is adopted:

$$N_w = N_0(1 - cos\,\theta) exp\{f(P)\} \Delta T_{sup}^{A\Delta T_{sup}+B} \tag{15}$$

Where $N_0 = 1000\, site/m^2$, and $cos\,\theta$ represents the effect of the contact angle. $f(P)$, A, B and $cos\,\theta$ can be calculated as follows:

$$f(P) = 26.006 - 3.678\, exp(-2P) - 21.907\, exp(-\frac{P}{24.065}) \tag{16}$$

$$A = -0.0002P^2 + 0.0108P + 0.0119 \tag{17}$$

$$B = 0.122P + 1.988 \tag{18}$$

$$1 - cos\,\theta = (1 - cos\,\theta_0)(\frac{T_c - T_{sat}}{T_c - T_0}^{\gamma}) \tag{19}$$

This model describes the effects of wall superheat, pressure, and contact angle on the nucleation site density. The validation results reported by Li et al. demonstrated that the model predictions are in good agreement with a wide range of published experimental data, as shown in Fig.2[24].

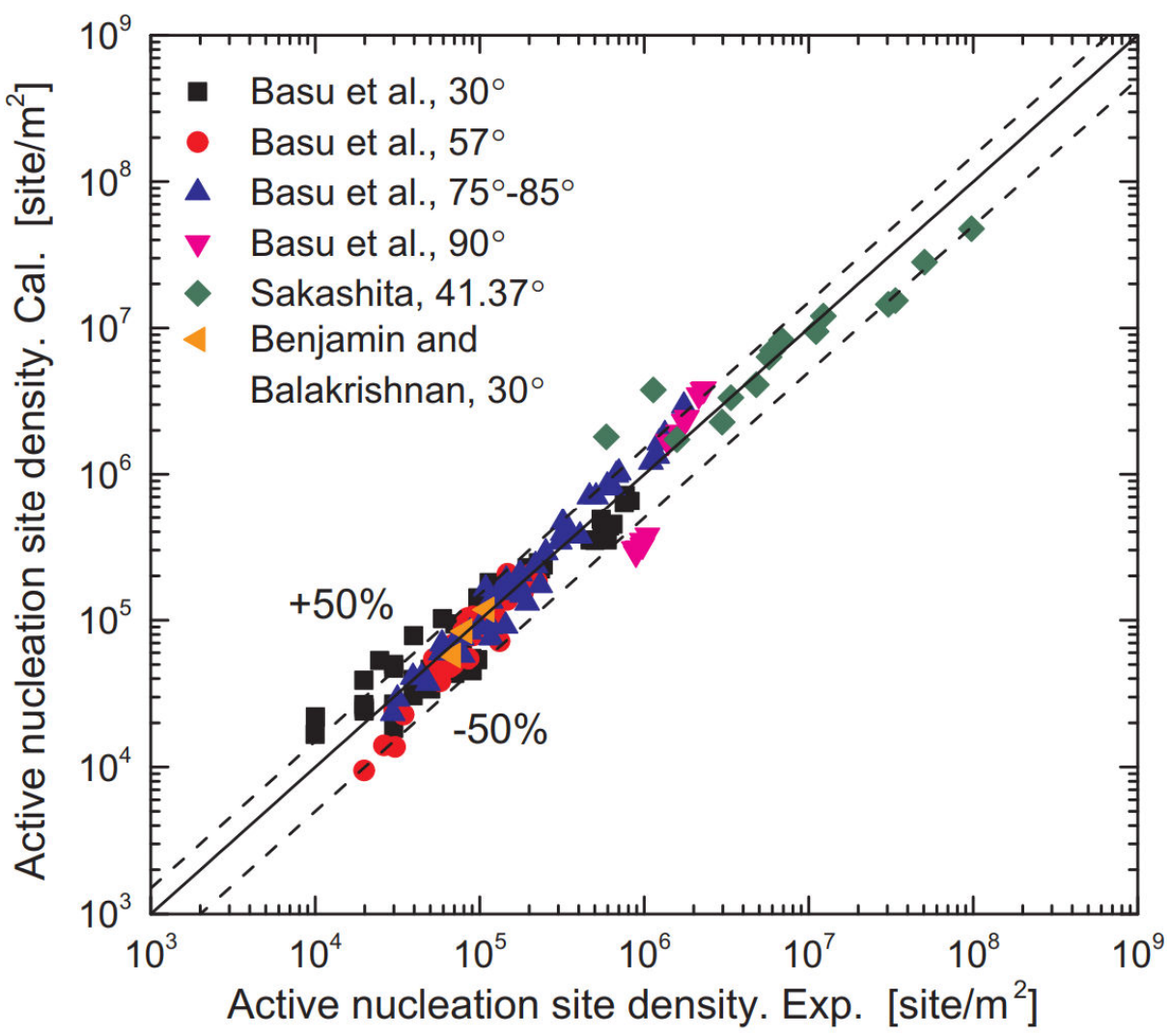

Fig.2 Data from the work by Li et al.[24]: statistical analysis of the new model with Basu, Sakashita and Benjamin's experiments data

Within this modeling framework, the number of activated nucleation sites on the superheated wall can be predicted based on the local thermal conditions. A key issue that then arises is how to generate a physically realistic and numerically stable spatial distribution of these nucleation sites for a given site density. Based on the studies of Sara Youssoufi et al.[22] and H. Faure et al.[29], the spatial distribution of nucleation sites on metallic surfaces can be approximated using a Halton low-discrepancy sequence.

The Halton-sequence is a deterministic low-discrepancy sequence that generates quasi-random points with relatively uniform coverage of a prescribed domain. Compared with purely random sampling, it can reduce excessive local clustering and large empty regions for a finite number of points, and the generated point set is exactly reproducible. In the present work, the Halton-sequence is used as a numerical allocation strategy to distribute the active nucleation sites predicted by the nucleation site density model over the heated wall. Similar low-discrepancy or Halton-sequence based distributions have also been adopted in previous high-fidelity boiling simulations to obtain reproducible and relatively uniform nucleation-site placement. Nevertheless, this treatment should not be interpreted as an exact representation of the microscopic cavity distribution on a real surface, which depends on surface roughness, wettability, defects, contamination, and manufacturing processes. Instead, the Halton based distribution provides a controlled and reproducible approximation for multi-site nucleation when detailed surface morphology is unavailable. Its main purpose is to reduce the arbitrariness associated with manually prescribed nucleation sites and to avoid the excessive clustering that may occur in purely random distributions.

Accordingly, in the present work, the predicted nucleation sites are mapped onto the superheated surface using a Halton-sequence. First, a two-dimensional (2D) Halton point set is generated and scaled according to the geometric dimensions of the heated wall. These points are then projected onto the nearest wall cells, and the corresponding cell centers are identified as nucleation sites. To initialize and update vapor embryos during the phase field/VOF iterations, a hemispherical mask is constructed around each nucleation site. The initial nucleus radius $r_0$ is typically set to 1 − 2 local grid spacings to ensure consistency between the nucleus size and numerical resolution while avoiding numerical instability. In addition, to reduce computational cost, the nucleation site density and spatial distribution are updated every ten iterations in the present simulations.

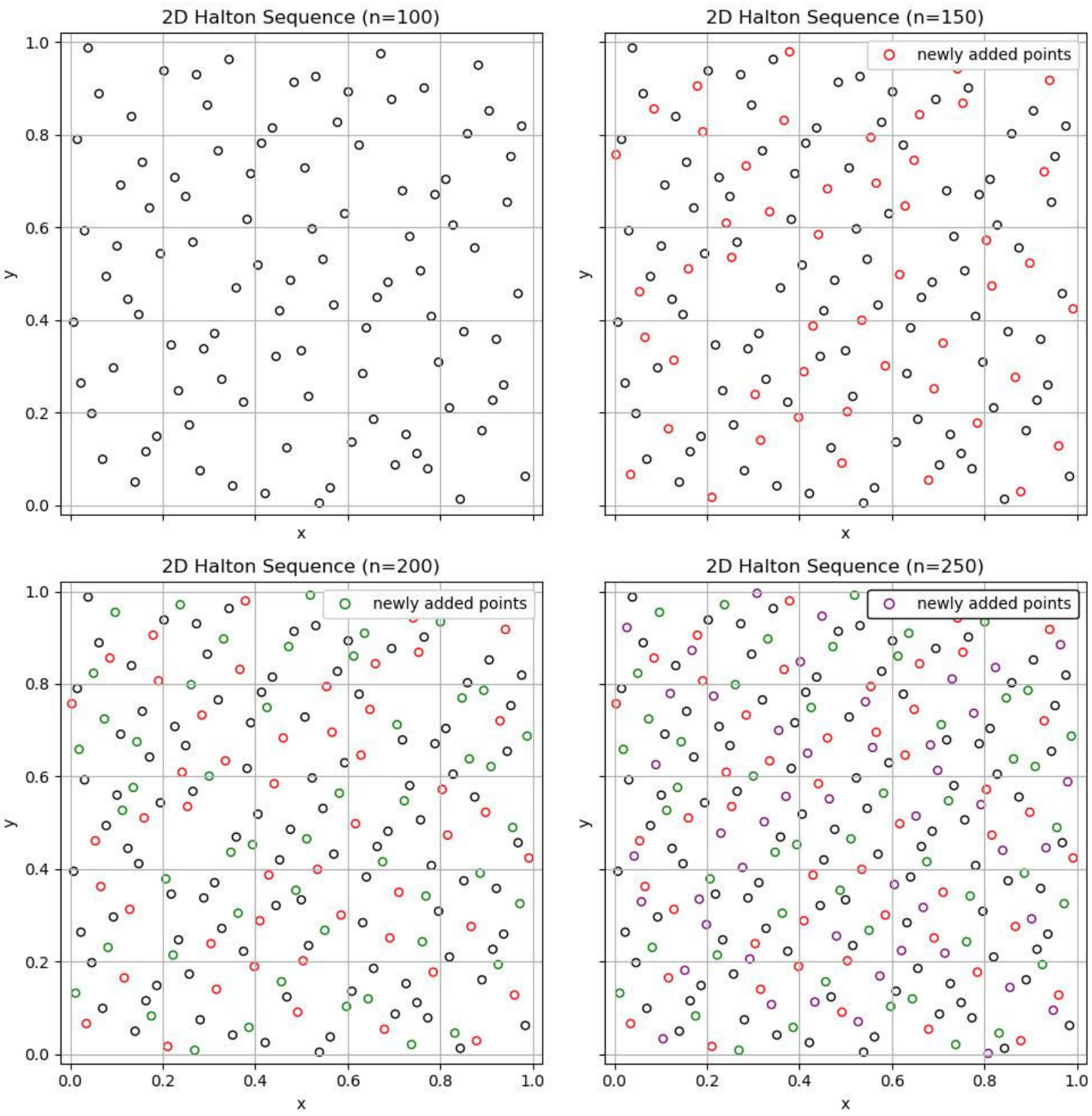

Fig.3 Schematic of the spatial distribution of Halton-sequence nucleation sites on a 2D rectangular surface for different numbers of active sites

Compared with conventional random distributions, the Halton based approach offers several advantages: (1) improved reproducibility and consistency, avoiding frequent relocation of nucleation sites and thereby reducing non-physical fluctuations; (2) a more uniform spatial distribution due to its low-discrepancy property, which minimizes overlap and excessive clustering of nucleation sites; and (3) enhanced numerical stability and computational efficiency for a given number of sites, owing to reduced overlap checks and resampling requirements. An illustrative example of the nucleation site number and spatial distribution on a rectangular heated surface as a function of wall superheat is presented in Fig.3. The detailed implementation procedure for the wall-nucleation update and Halton-sequence based nucleation mask construction is provided in the Supplementary Material.

## E. Numerical method

The governing equations were discretized using the finite volume method on a collocated mesh within a C++ based open-source finite-volume framework OpenFOAM-v2506. In the present simulations, bubble growth and vapor distribution result from the coupled phase-change, nucleation, flow, and heat transfer processes. A conjugate heat transfer strategy was employed to resolve the coupled fluid-solid thermal transport. At each iteration, the governing equations were solved sequentially in the fluid and solid regions. In the solid region, only heat conduction was considered, whereas in the fluid region a transient PIMPLE loop was used to advance the momentum and energy equations. The pressure-velocity coupling was handled using the PIMPLE algorithm for transient incompressible two phase flows.

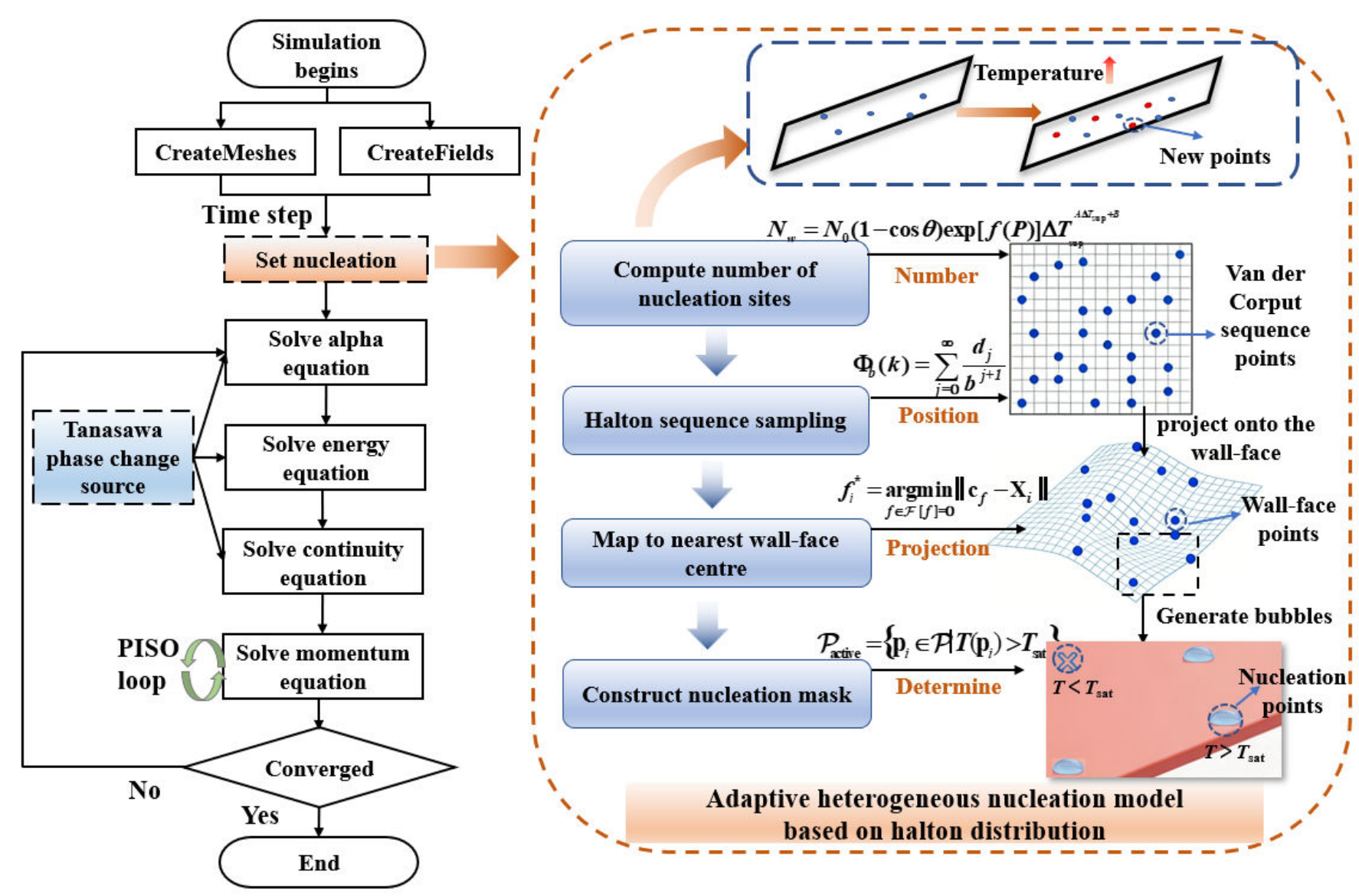

Fig.4 The flow diagram of the numerical solution framework for the boiling process

Table.3 summarizes the key simulation parameters and numerical controls used in the present simulations. Time integration was performed using a first-order implicit Euler scheme. The time step was adaptively controlled with an initial value of $\Delta t = 1 \times 10^{-6}$ s and a maximum value of $\Delta t = 1 \times 10^{-4}$ s, while the maximum fluid Courant number was limited to $Co_{max} = 0.1$. For the conjugate solid region, the maximum diffusion number was limited to $Di_{max} = 10$. Spatial gradients were evaluated with a Gauss linear scheme, while diffusive terms were discretized with a Gauss linear corrected Laplacian to account for mesh non-orthogonality. The wall normal temperature gradient required for heat flux evaluation employed a corrected surface normal gradient scheme. For the convective terms in the momentum and energy equations, bounded second-order accurate schemes were adopted to ensure stability in the presence of strong interfacial property jumps. Specifically, the momentum convection term was discretized using a Gauss limitedLinearV scheme, while the temperature convection term was discretized using a Gauss limitedLinear scheme. Viscous stress divergence terms were treated with a Gauss linear scheme. The linear solver tolerances for the volume-fraction, pressure, and temperature equations were set to $10^{-8}$, $10^{-9}$, and $10^{-9}$, respectively. Interface capturing was achieved with a geometric VOF approach based on the isoAdvector method. The phase fraction advection was discretized using a bounded high resolution van Leer scheme. The interface geometry required by isoAdvector was reconstructed using a gradient-based approach, which provides a robust estimate of the interface normal from $\nabla\alpha$ and supports piecewise linear interface construction within interfacial cells. During the isoAdvector update, the boundedness of the liquid volume fraction was monitored through $min(\alpha_l)$ and $max(\alpha_l) - 1$ with a numerical tolerance of $10^{-13}$. Conservative bounding correction was applied if undershoots or overshoots were detected. The stability of interface tracking was further checked using the Courant-number control and the continuity-error output during the pressure-correction loop. A mesh independence study was carried out by monitoring the area-averaged heat flux transferred from the solid region to the fluid region under different mesh resolutions. The relative difference in the averaged heat flux was limited to approximately 1.5%, indicating satisfactory grid convergence for the present simulations.

Table.3 Summary of key simulation parameters and numerical controls

| Category | Setting |
|---|---|
| Reynolds number | 1000~5000 |
| Bottom temperature | 340~420 K |
| Averaging window | 0.4~1.0 s |
| Initial time step | $1 \times 10^{-6}$ s |
| Maximum time step | $1 \times 10^{-4}$ s |
| Maximum fluid Courant number | 0.1 |
| Maximum solid diffusion number | 10 |
| linear solver tolerances for volume-fraction | $1 \times 10^{-8}$ |
| linear solver tolerances for pressure | $1 \times 10^{-9}$ |
| linear solver tolerances for temperature | $1 \times 10^{-9}$ |

## III. NUMERICAL FRAMEWORK VALIDATION

To comprehensively validate the mathematical models and numerical framework employed in the present study, two representative benchmark cases were considered, namely the 2D Stefan problem and bubble growth in a microchannel. These validation cases confirm the capability of the solver in predicting fluid flow, heat transfer, and phase change processes. In addition, selected results of saturated flow boiling in a rectangular microchannel obtained using the present solver were statistically compared with a widely used empirical correlation to further assess the reliability of the model.

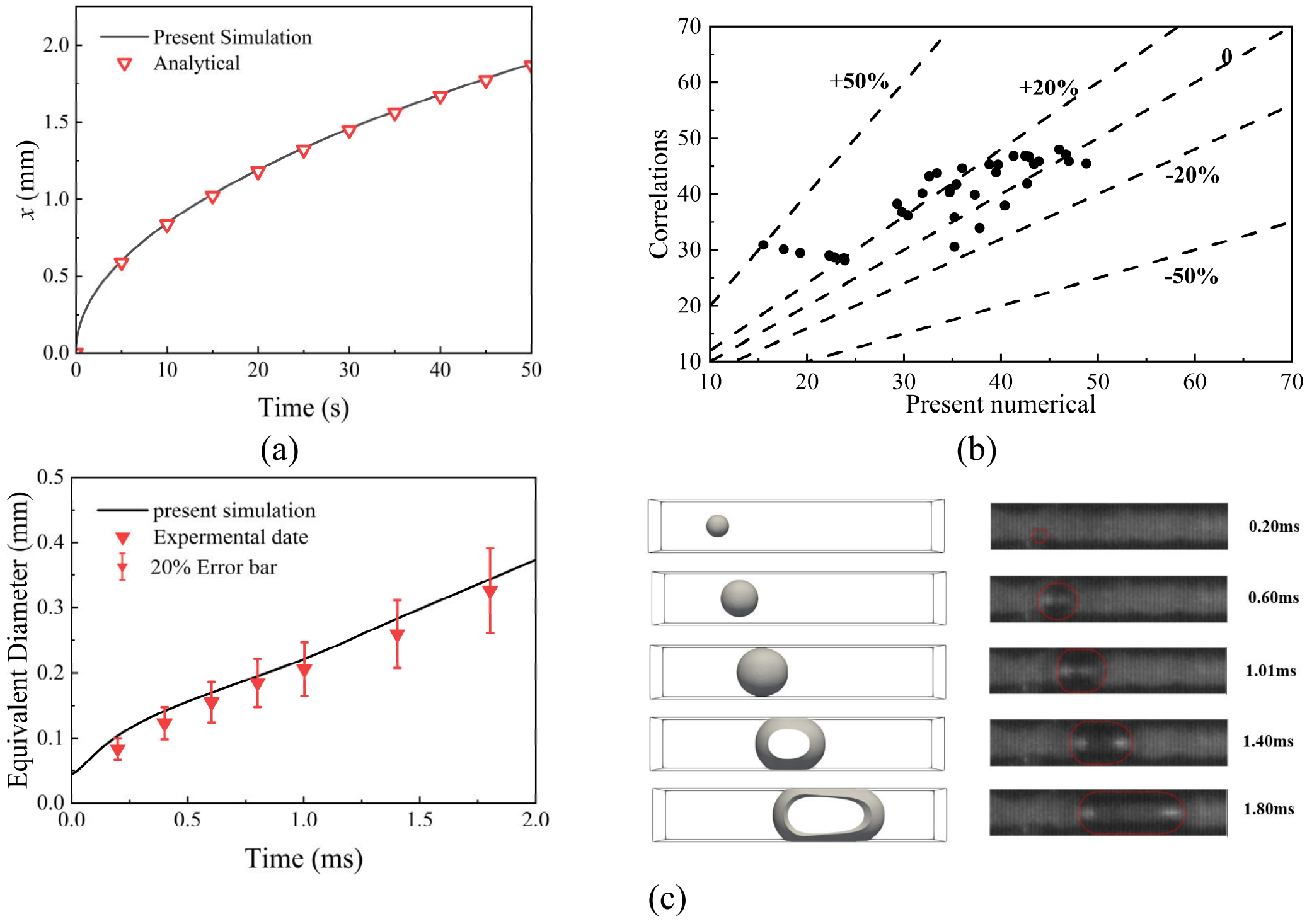

Fig.5 Validation of the 3D finite volume based numerical framework: (a) validation against the 2D Stefan problem; (b) comparison of the numerical results for microscale saturated flow boiling with commonly used empirical correlations; (c) validation of single bubble growth in a microchannel: comparison of bubble diameter evolution between numerical simulation and experimental measurements

The Stefan problem[30] is a standard benchmark for boiling and phase change simulations, and its analytical solution is widely used to evaluate the accuracy of numerical methods. In the present study, the 2D Stefan problem was simulated, and Fig.5-(a) compares the numerical results with the analytical solution. It can be seen that the numerical results are in good agreement with the analytical solution, confirming the accuracy of the present numerical treatment of phase change.

Fig.5-(b) summarizes numerical results for saturated flow boiling in a microchannel and compares them with the empirical correlation proposed by Li and Wu[31]. It can be seen that all the data fall within the ±50% confidence band, and approximately 68% of the data lie within the ±20% error range, indicating good reliability of the present numerical framework.

Fig.5-(c) presents the validation against the microchannel single bubble growth experiment reported by Mukherjee[32]. The predicted bubble diameter evolution and interface morphology are in good agreement with the experimental results. Overall, the simulation captures the key dynamics of bubble growth, thereby supporting the validity of the present numerical framework for flow boiling simulations.

It should be noted that the present validation is mainly intended to verify the essential components of the numerical framework, including the phase-change treatment, single-bubble growth dynamics, and the overall level of flow boiling heat transfer. A direct validation of the predicted dryout topology and CHF trend for the exact rectangular mini-channel configuration considered here remains difficult because directly comparable experimental data with the same working fluid, geometry, conjugate substrate, heating mode, surface condition, and inlet condition are not currently available. Therefore, the *Re*-dependent transition pathways identified in this study should be interpreted as numerical observations within the present configuration and operating range.

# IV. RESULTS

This section examines the effects of 5 different inlet *Re* ($Re = \frac{\mathbf{u}L}{\nu}$) on the boiling mode and heat transfer characteristics in rectangular mini-channel. The selected cases represent two distinct inlet flow conditions, namely relatively weak liquid replenishment (*Re*=1000~2000) and stronger liquid replenishment (*Re*=3000~5000). For all cases, the bottom wall temperature of the solid substrate was varied from 340~420 K, which is sufficient to drive the boiling process from incipient nucleate boiling to film boiling. In this work, the contact angle was fixed at $90^{\circ}$, the ambient pressure was set to standard atmospheric pressure, and the thermal boundary conditions in the numerical simulations were prescribed as follows: the bottom superheated surface of the copper substrate was imposed as a constant temperature Dirichlet boundary $T_b$, while the initial temperature at the upper boundary, i.e., the conjugate heat transfer interface, was set to $T_b - 1$. Under these conditions, the inlet *Re* serves as the primary dimensionless control parameter in the present study. In the present work, the wall dryout fraction is defined as a continuous area-averaged vapor coverage, or dryness index over the conjugate heated surface:

$$f_{dryout} = \frac{1}{S}\int_{s}(1-\alpha_l)dS$$

where *S* is the area on the conjugate heated-wall patch and $\alpha_l$ is the corresponding local liquid volume fraction. Therefore, $f_{dryout} = 0$ corresponds to a fully liquid-wetted heated surface, whereas $f_{dryout} = 1$ corresponds to a fully vapor-covered heated surface. Intermediate values represent partial near-wall vapor occupation and partial loss of wall wetting. Since this definition uses the full local volume fraction information, it avoids introducing an arbitrary dry/wet threshold in the post-processing. For the parameter range considered in the present study, the system was observed to reach a stable quasi-thermal equilibrium state within 0.2 s, during which the fluctuation of the area-averaged temperature over the conjugate heated surface was less than 1% of $T_b$. Therefore, the time interval from 0.4 to 1.0 s was selected as the time-averaging window in this work. This averaging window is also appropriate for the heat transfer and wall dryness statistics because the local heat flux and Nusselt number are directly determined by the wall thermal state through the conjugate heat transfer process. Meanwhile, the wall dryout fraction is closely related to the wall nucleation activity, and the nucleation-site-density model adopted in this study is primarily controlled by the local wall superheat. All subsequent time-averaged quantities reported in the following sections were calculated using this averaging window. The present analysis focuses on the statistically stable quasi-steady boiling stage, in which the instantaneous vapor structures are used as representative visual evidence and the macroscopic heat transfer characteristics are evaluated from time-averaged quantities. It should be noted that this analysis is intended as a representative temporal-stability check for the selected averaging window, rather than a complete averaging-window convergence study for all

$Re$ and $T_b$ cases.

Based on this setup, the results and discussion are organized into two parts. Section A first identifies the dominant boiling modes and clarifies their corresponding transition pathways under different $Re$, with emphasis on how the competition between liquid replenishment, vapor growth, and interfacial reorganization gives rise to distinct transition behaviors. Section B then analyzes the associated heat transfer characteristics, including the evolution of the boiling curve, Nusselt number, and wall dryout behavior, in order to establish the link between boiling mode transition and overall thermal response.

**A. Boiling mode identification and transition processes**

a) Localized dryout dominated boiling transition mode ($Re$=1000~2000)

The low $Re$ cases ($Re$=1000, 2000) exhibit a common boiling transition pathway characterized by localized dryout development. All the snapshots shown in this subsection were taken at the same statistically stable instant, so that the observed differences can be attributed primarily to the variation in wall temperature rather than transient fluctuations. **A key feature of this mode is the pronounced axial differentiation of the boiling structure: the upstream part of the channel remains actively boiling, whereas the downstream part progressively loses liquid supply and evolves into a locally dried region.** Therefore, in the present simulations, the transition is mainly associated with the progressive expansion of downstream liquid starvation and local dryout, rather than with global vapor reorganization over the entire channel.

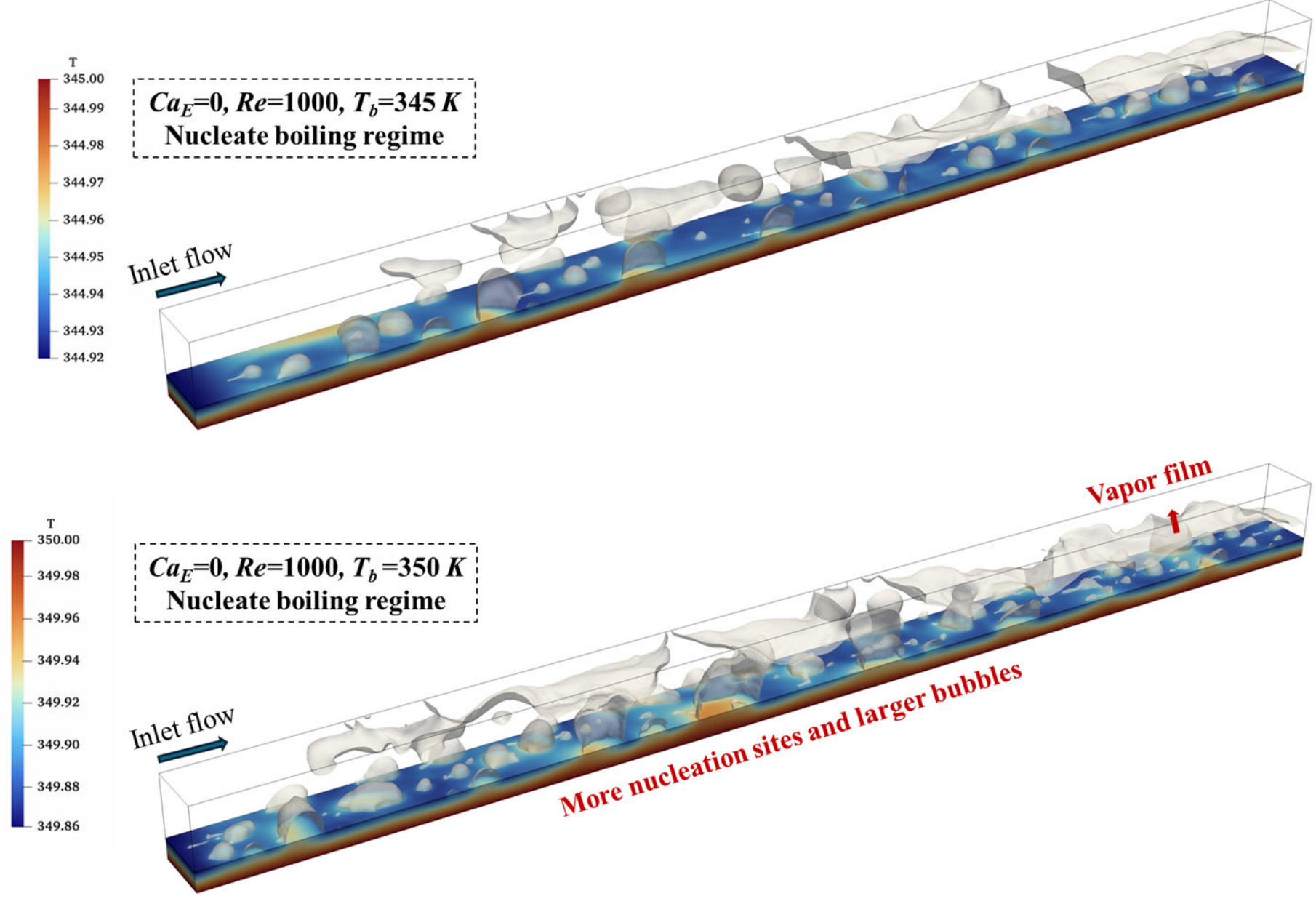

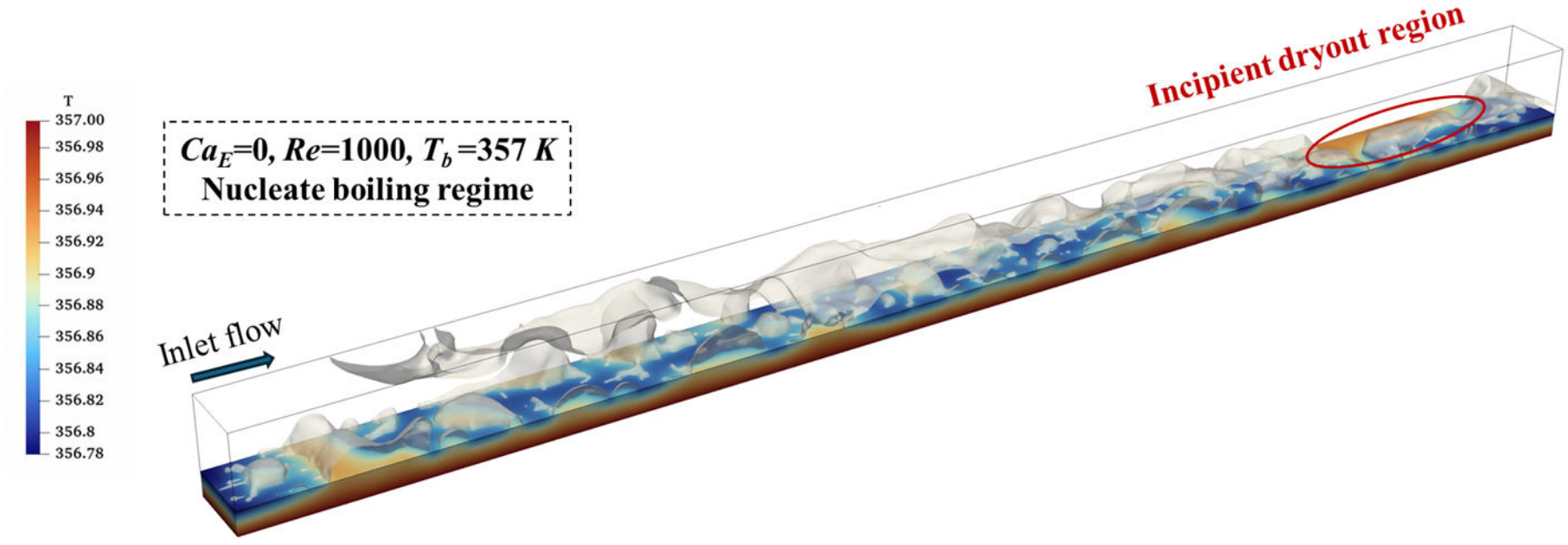

Fig.6 Evolution of the localized dryout dominated boiling transition mode for *Re*=1000: from nucleate boiling to incipient downstream dryout

The following results show the evolution of the localized dryout dominated boiling transition mode at *Re*=1000. Fig.6 shows the process from nucleate boiling to incipient downstream dryout. For *Re*=1000, the boiling process initially follows a typical nucleate boiling route. At relatively low wall temperatures ($T_b$=345~355 K), discrete vapor bubbles are generated from multiple nucleation sites and grow gradually with increasing wall superheat. Meanwhile, bubble coalescence becomes more pronounced and vapor accumulates near the upper part of the channel ($T_b$=350 K). As $T_b$ increases further, the amount of liquid remaining in the downstream part of the channel decreases noticeably, and an incipient dryout region first appears near the outlet at $T_b$≈357 K. This indicates that the local liquid replenishment is no longer sufficient to compensate for the intensified evaporation in the upstream and middle parts of the channel.

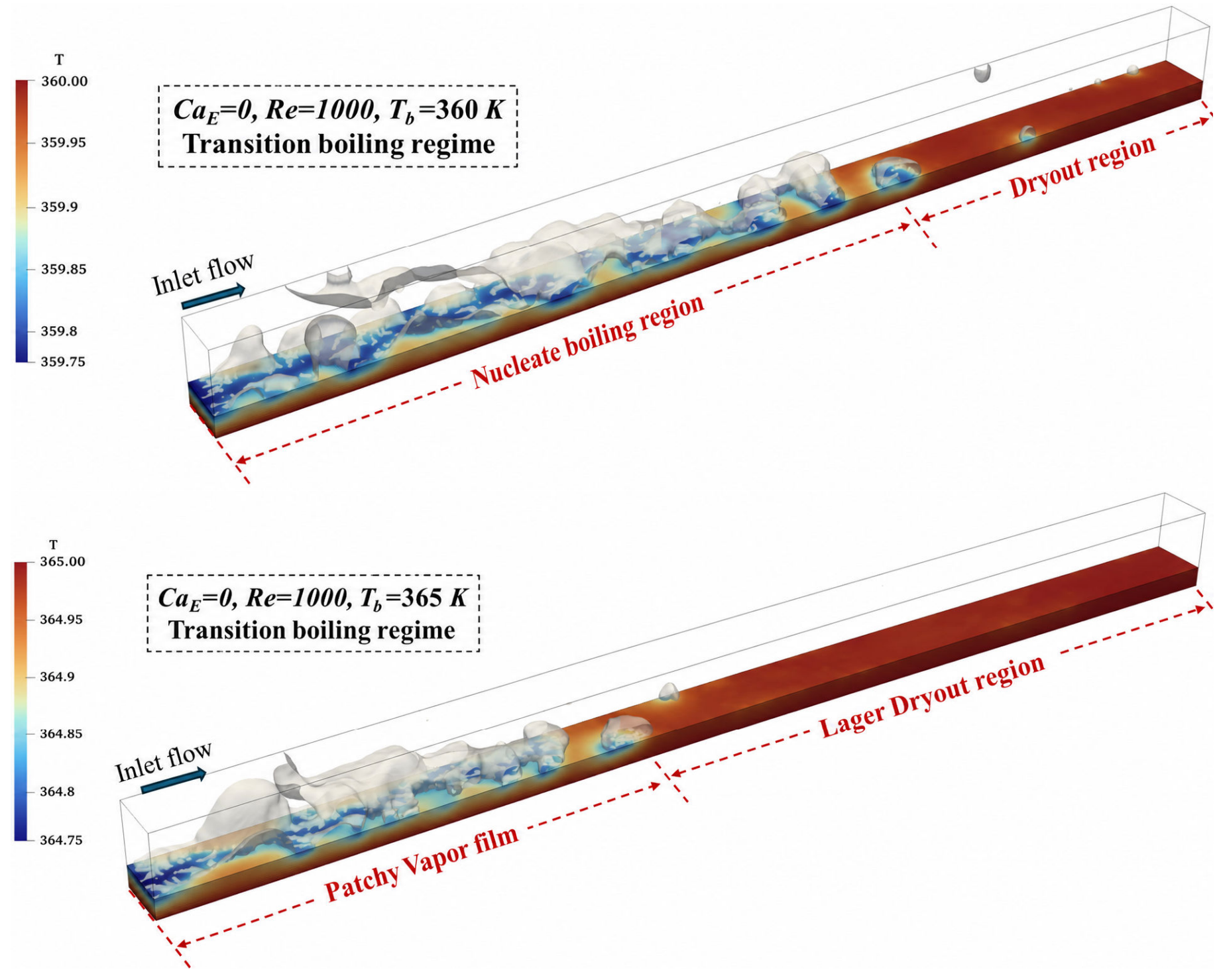

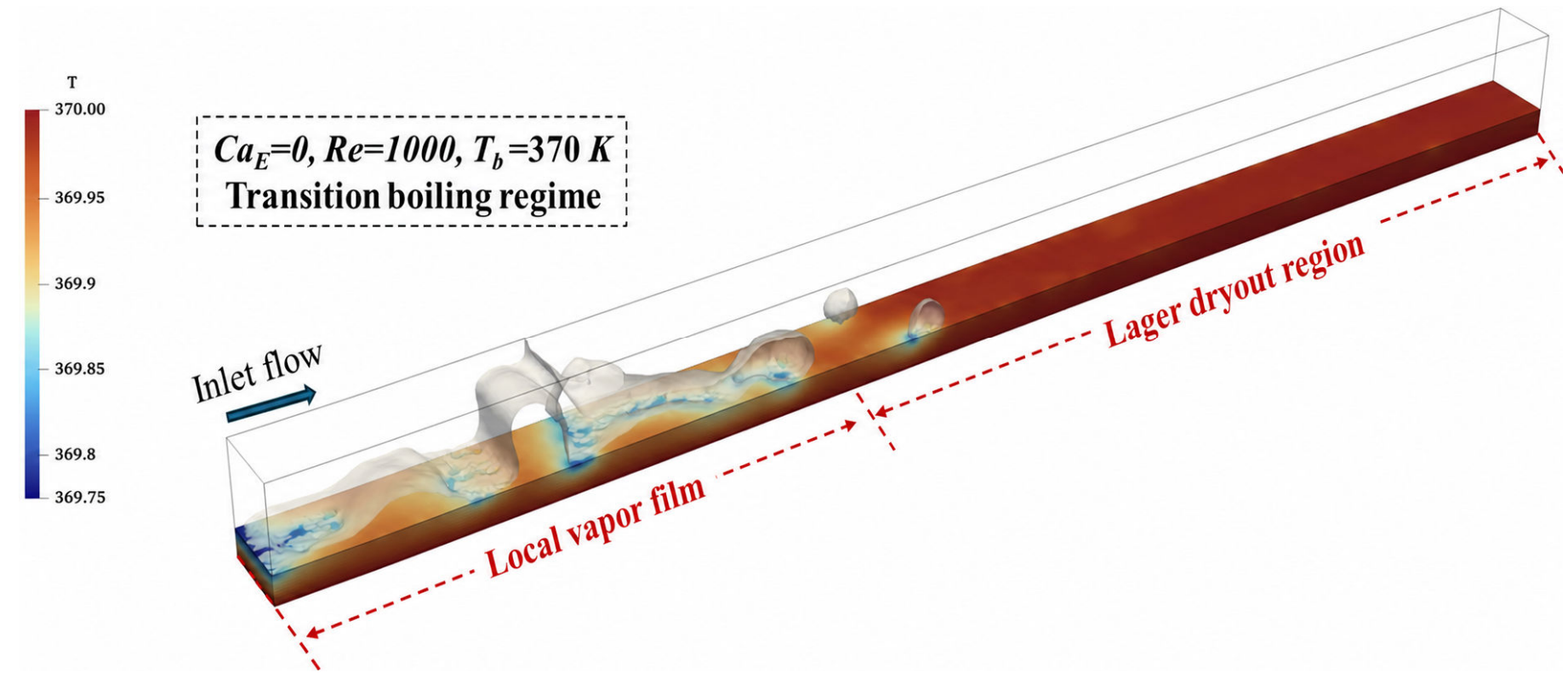

Fig.7 Localized dryout dominated boiling transition mode for *Re*=1000: transition boiling stage

When $T_b$ is further raised to $T_b$=360~370 K, the system enters the transition boiling regime, which is shown in Fig.7. In this stage, a clear spatial separation develops along the streamwise direction. The upstream region still exhibits active nucleate boiling or patchy/local vapor film structures, while the downstream half of the channel becomes dominated by an extended dryout region. Physically, this behavior suggests that a large fraction of the incoming liquid is vaporized in the front part of the channel, leaving insufficient liquid to rewet the downstream wall. As a result, the dryout region expands progressively upstream with increasing wall temperature, while local vapor film structures begin to appear in the upstream boiling zone as well.

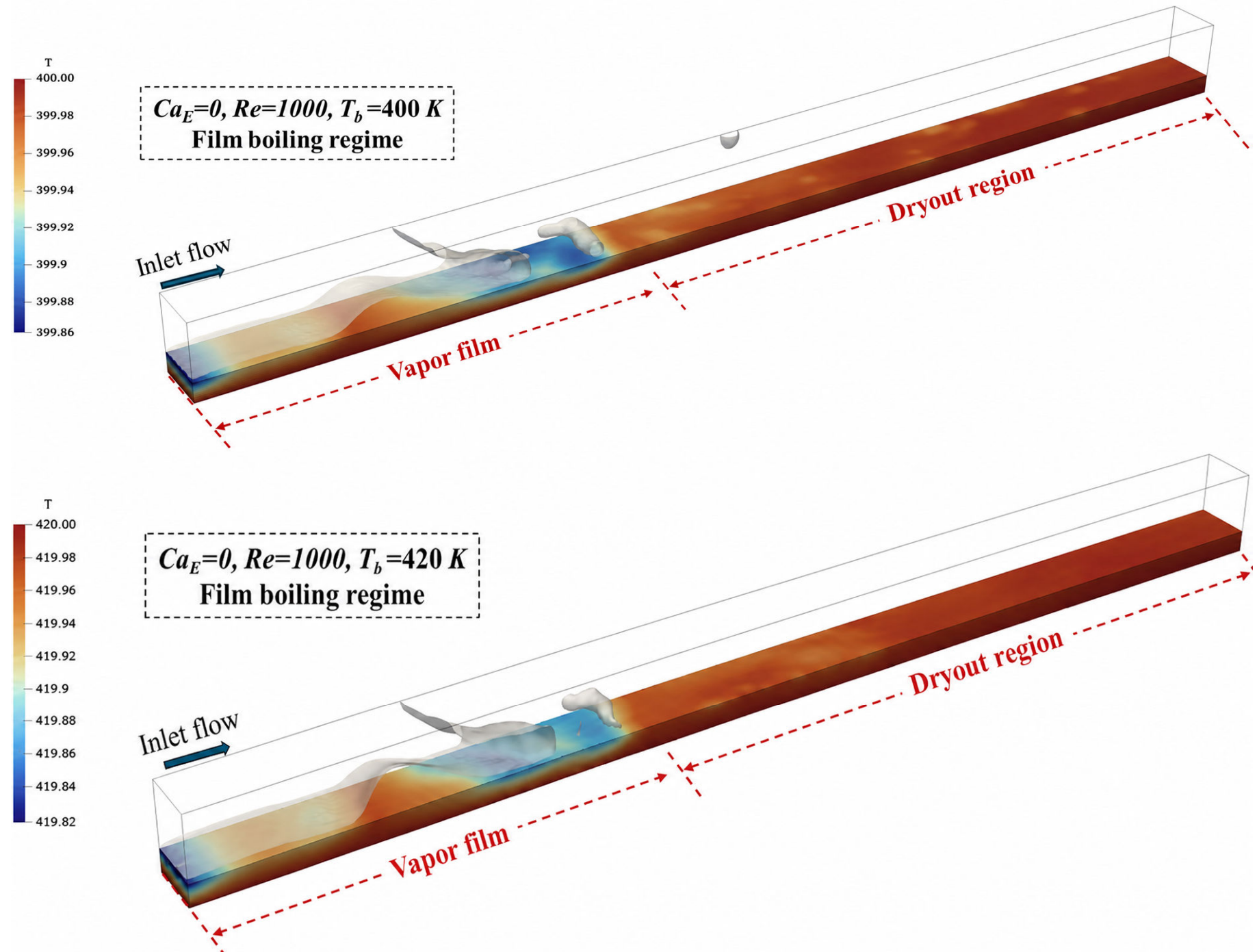

Fig.8 High temperature stage of the localized dryout dominated transition mode for *Re*=1000: coexistence of an upstream vapor film region and a downstream dryout region

At sufficiently high wall temperatures ($T_b$=400~420 K), the channel evolves into a relatively stable film boiling dominated configuration, in which an upstream vapor film region coexists with a downstream dryout region. The relative proportions of these two regions vary only weakly with further increase in $T_b$, indicating that, under a fixed inlet liquid supply, most of the liquid is already consumed and vaporized in the upstream part, whereas the downstream wall remains in a liquid starved state.

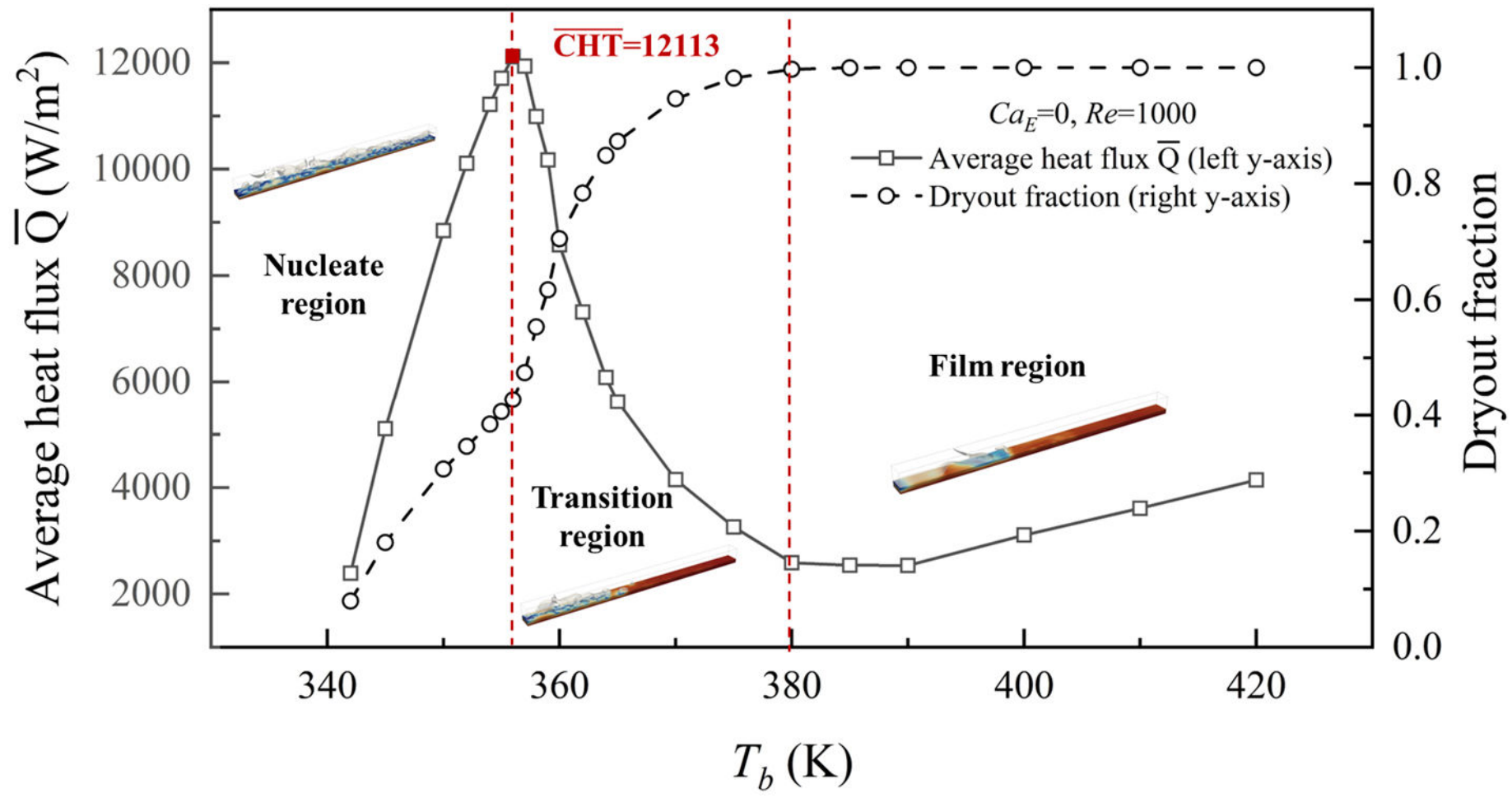

Fig.9 Time-averaged heat flux and wall dryout fraction at conjugate superheated surface for $Re$=1000

The above transition process is directly reflected in Fig.9, which reflects the relationship between the time-averaged heat flux and the dryout fraction at the superheated conjugate surface. For $Re$=1000, the heat flux first rises rapidly with increasing wall temperature and reaches a maximum of approximately 12.1 kW/m$^2$ at $T_b$≈356 K, corresponding to the most efficient nucleate boiling stage. We define the maximum value of the time-averaged spatially averaged heat flux over the conjugate superheated surface as the time-averaged critical heat flux, $\overline{\mathrm{CHF}}$. Although dryout has already started to emerge around this point, it has not yet become dominant over the entire heating surface. Beyond this peak, however, the dryout fraction increases sharply, accompanied by a pronounced drop in the heat flux. This close correspondence indicates that the deterioration of boiling heat transfer is strongly associated with the rapid expansion of wall dryout. At even higher wall temperatures, the heat flux exhibits a slight recovery, but the dryout fraction remains close to unity, indicating that this recovery is mainly caused by the larger imposed temperature difference rather than by any restoration of effective wall wetting.

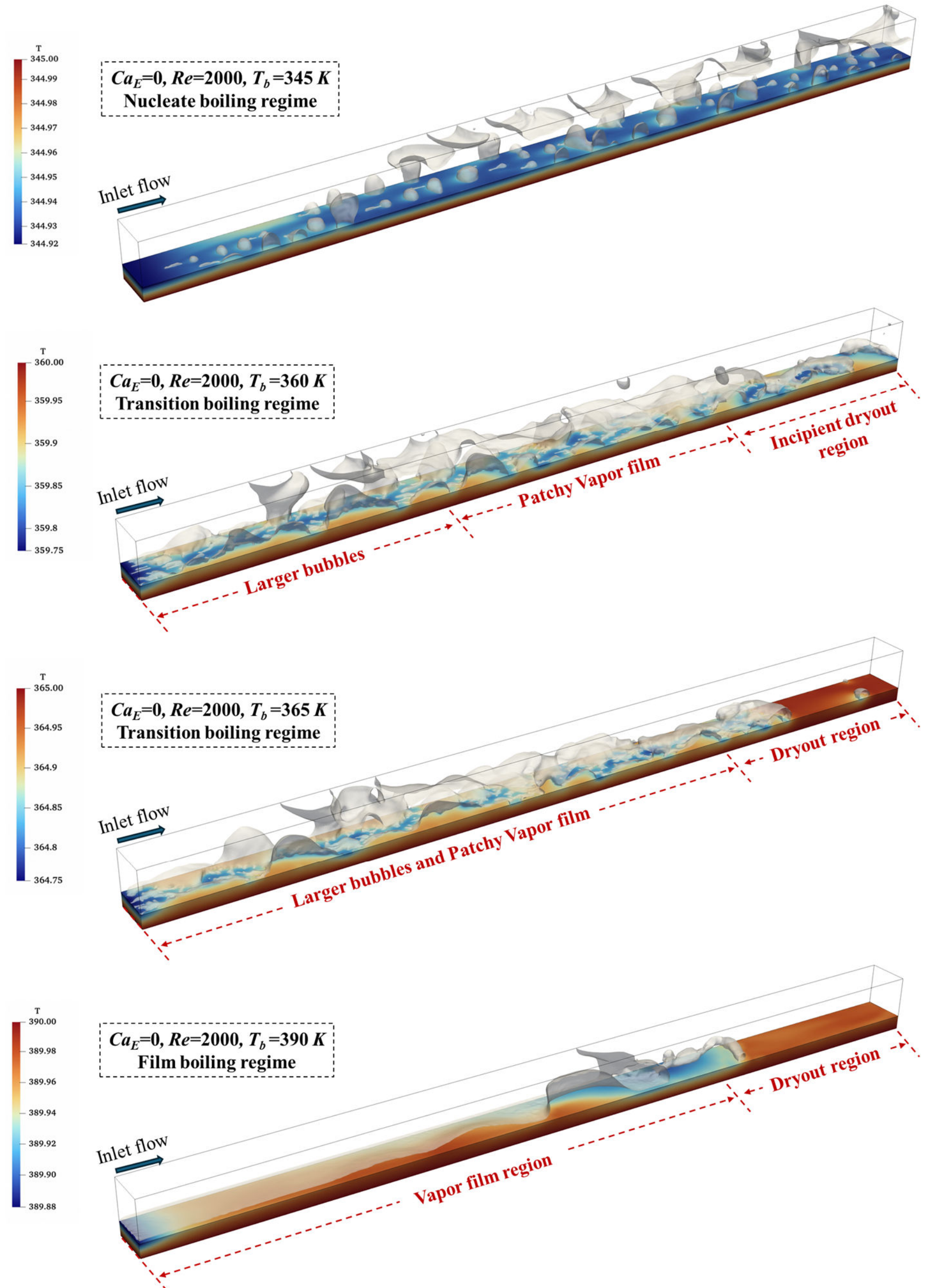

Fig.10 Localized dryout dominated boiling transition mode for *Re*=2000

Fig.10 shows the transition pathway at *Re*=2000. Increasing the inlet *Re* to 2000 does not alter the basic transition pathway, but it significantly affects its progression. The overall evolution still follows the sequence of bubble growth and coalescence, patchy vapor film formation, downstream liquid starvation, and local dryout expansion. However, owing to the stronger inlet liquid replenishment, the onset of severe dryout is delayed and the dryout region remains smaller than that in the *Re*=1000 case at comparable $T_b$.

This behavior is also reflected by the corresponding thermal response shown in Fig.11: the peak heat flux (11.69 kW/m$^2$ at $T_b$≈358 K) occurs at a slightly higher $T_b$, and the heat flux deterioration in the transition and high temperature regimes is less severe. Therefore, *Re*=1000 and 2000 belong to the same localized dryout dominated boiling transition family. The increase in *Re* mainly delays the onset of dryout controlled deterioration and weakens its severity.

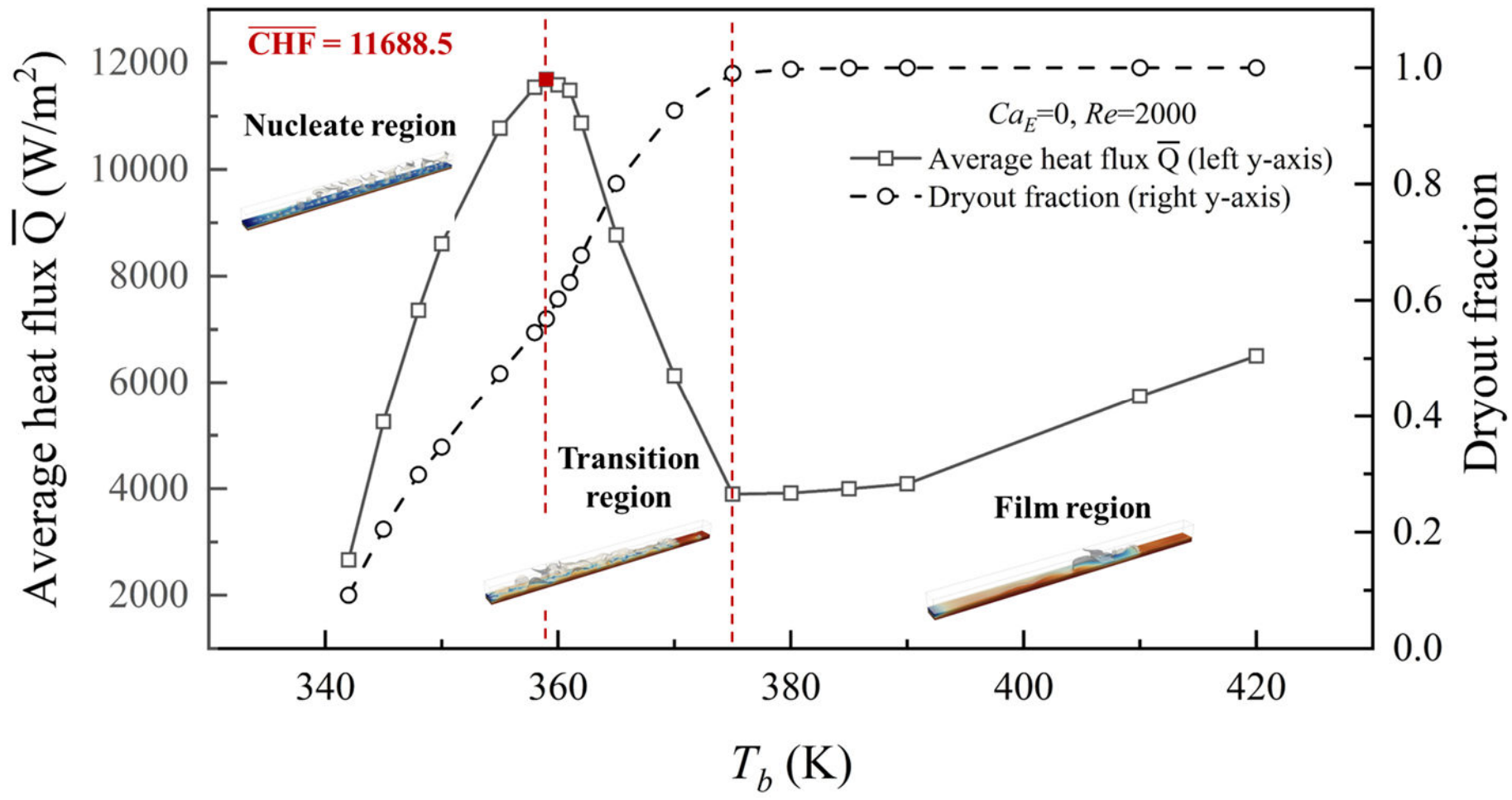

Fig.11 Time-averaged thermal response and wall dryout fraction at conjugate superheated surface for *Re*=2000

b) Convection dominated elongated vapor boiling transition mode (*Re*=3000~5000)

The higher *Re* cases (*Re*=3000~5000) exhibit a distinct boiling transition pathway characterized by convective vapor stretching and reorganization. In contrast to the low *Re* cases, where **the transition is triggered primarily by downstream liquid starvation and local dryout expansion, the present mode originates from the convective elongation of vapor structures**. Under stronger inlet, vapor bubbles are continuously stretched by the mainstream and evolve into elongated vapor slugs, which subsequently coalesce and form localized vapor films along the heated wall. Therefore, the transition process in this regime is mainly characterized by vapor-structure reorganization rather than by the direct upstream to downstream expansion of a dryout front.

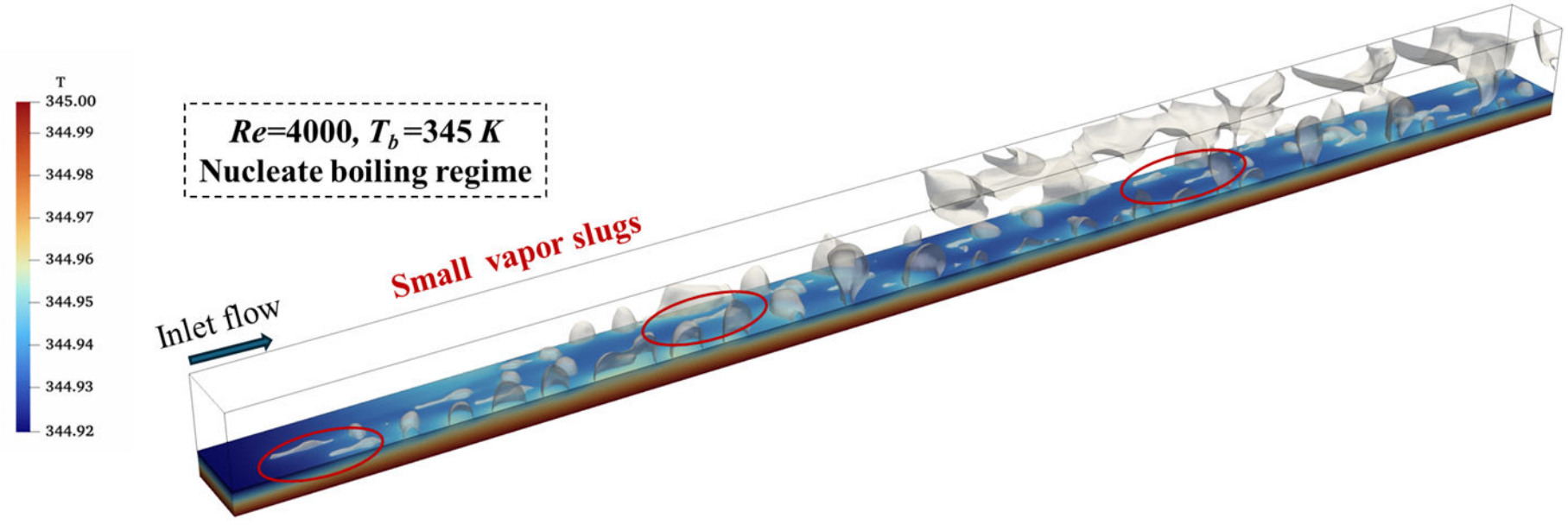

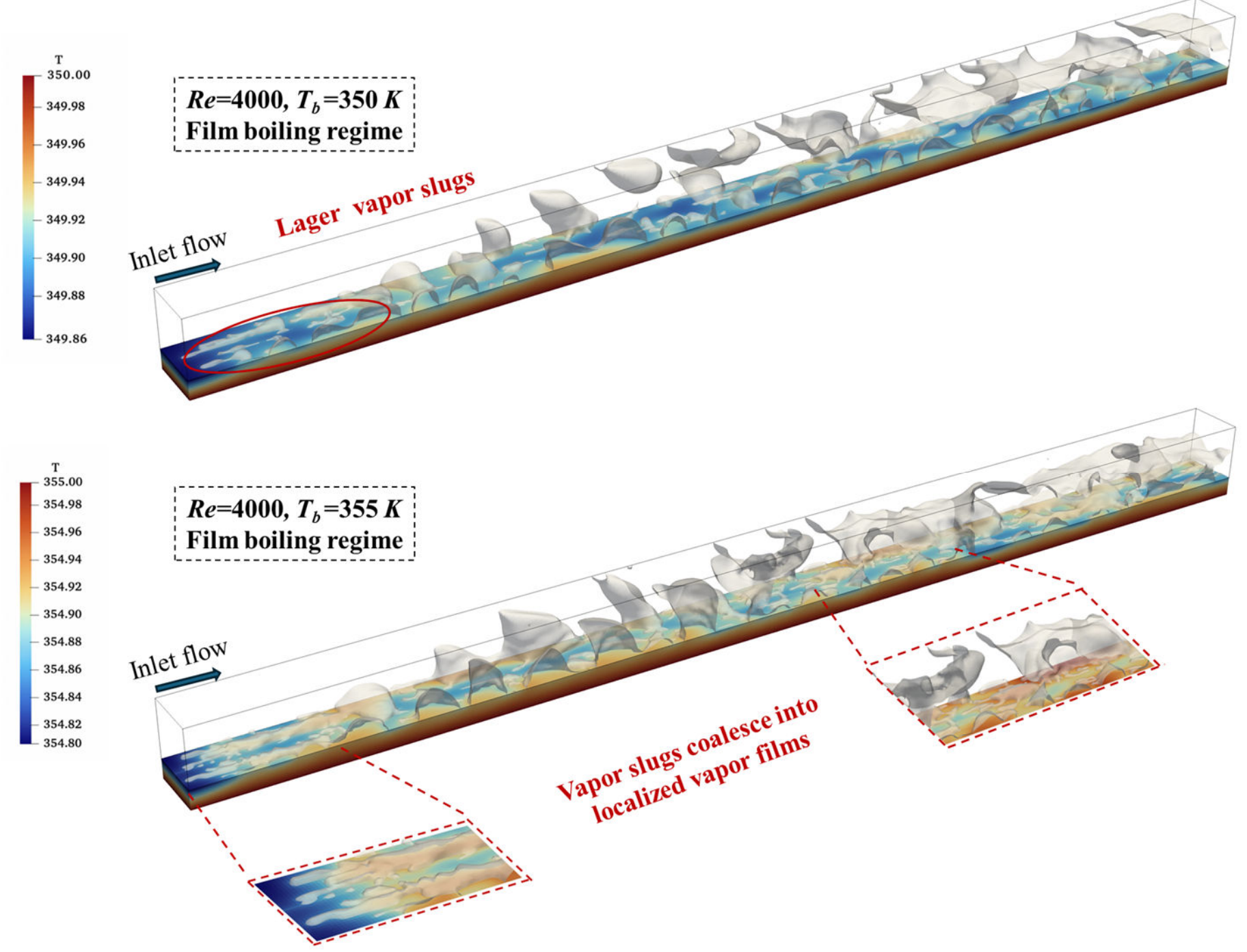

Fig.12 Early stage evolution of the convection dominated elongated vapor transition mode for *Re*=4000: from elongated vapor slugs to localized vapor films

The evolution of this mode is illustrated in detail using the *Re*=4000 case. At relatively low wall temperatures ($T_b$=345~355 K), the flow remains in the nucleate boiling stage, but the vapor structures already differ markedly from those observed at low *Re*. As shown in Fig.12, discrete bubbles are rapidly elongated by the incoming flow and form small vapor slugs even at $T_b$=345 K. With increasing wall temperature, these vapor slugs become larger and more continuous, and by $T_b$=355 K they start to coalesce into localized vapor films. This indicates that, under strong convective transport, the early boiling transition is controlled not by liquid depletion in the downstream region, but by the streamwise stretching and merging of vapor structures.

As shown in Fig.13, when the wall temperature is further increased to $T_b$=360~365 K, the system enters the transition boiling stage. In this regime, the localized vapor films expand significantly along the streamwise direction, while large vapor structures appear in the downstream region. Such behavior suggests that the residual liquid film beneath the elongated vapor structures becomes increasingly unstable under the combined effects of evaporation and shear, leading to local rupture and the formation of large downstream vapor pockets. As a result, the channel no longer exhibits the clear upstream boiling/downstream dryout separation observed in the low *Re* cases. Instead, the dominant feature becomes the progressive extension of localized vapor films over the heated surface, accompanied by the coexistence of large vapor structures in the downstream part of the channel.

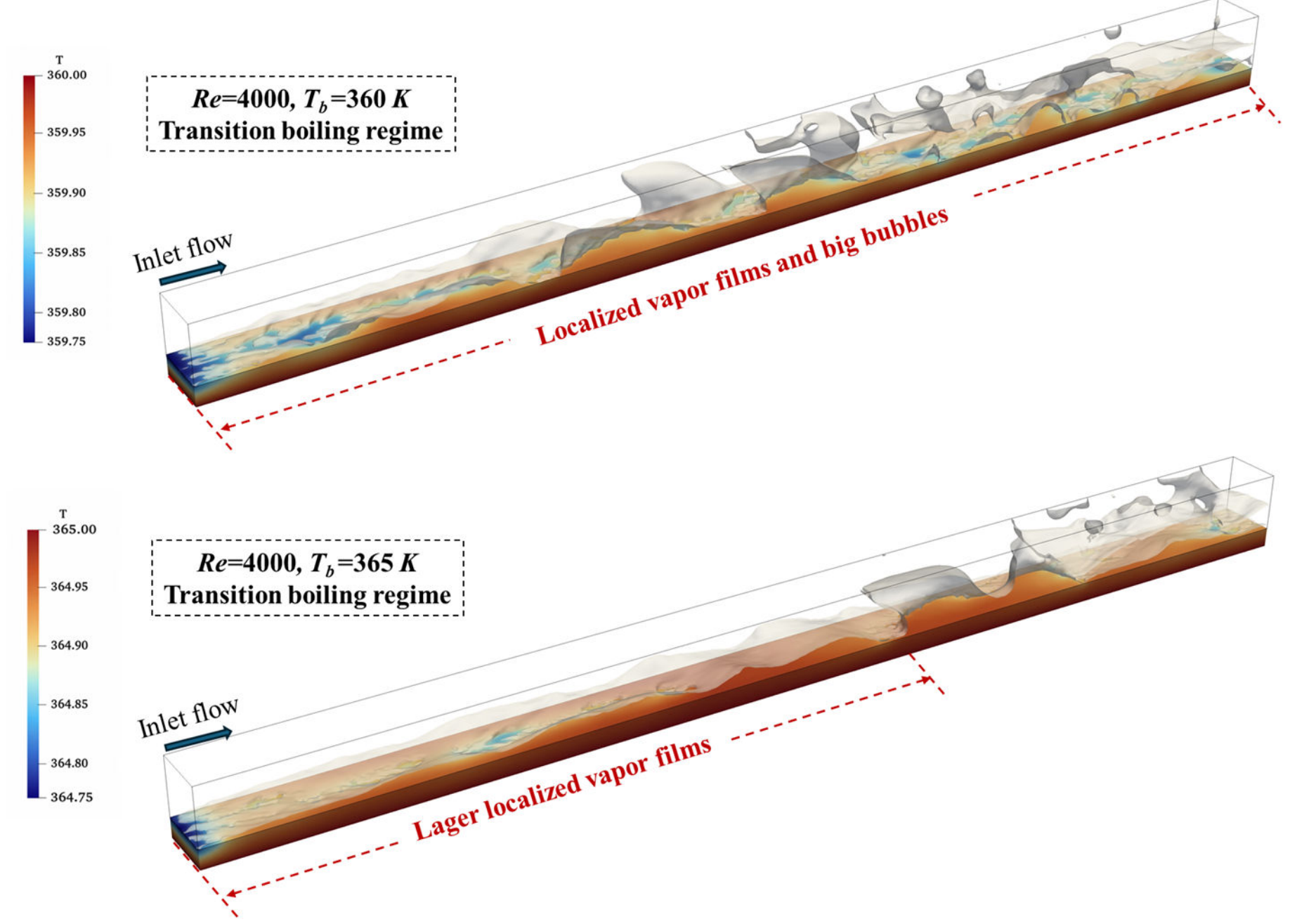

Fig.13 Transition boiling stage of the convection dominated elongated vapor mode for *Re*=4000: extension of localized vapor films and downstream large vapor structures

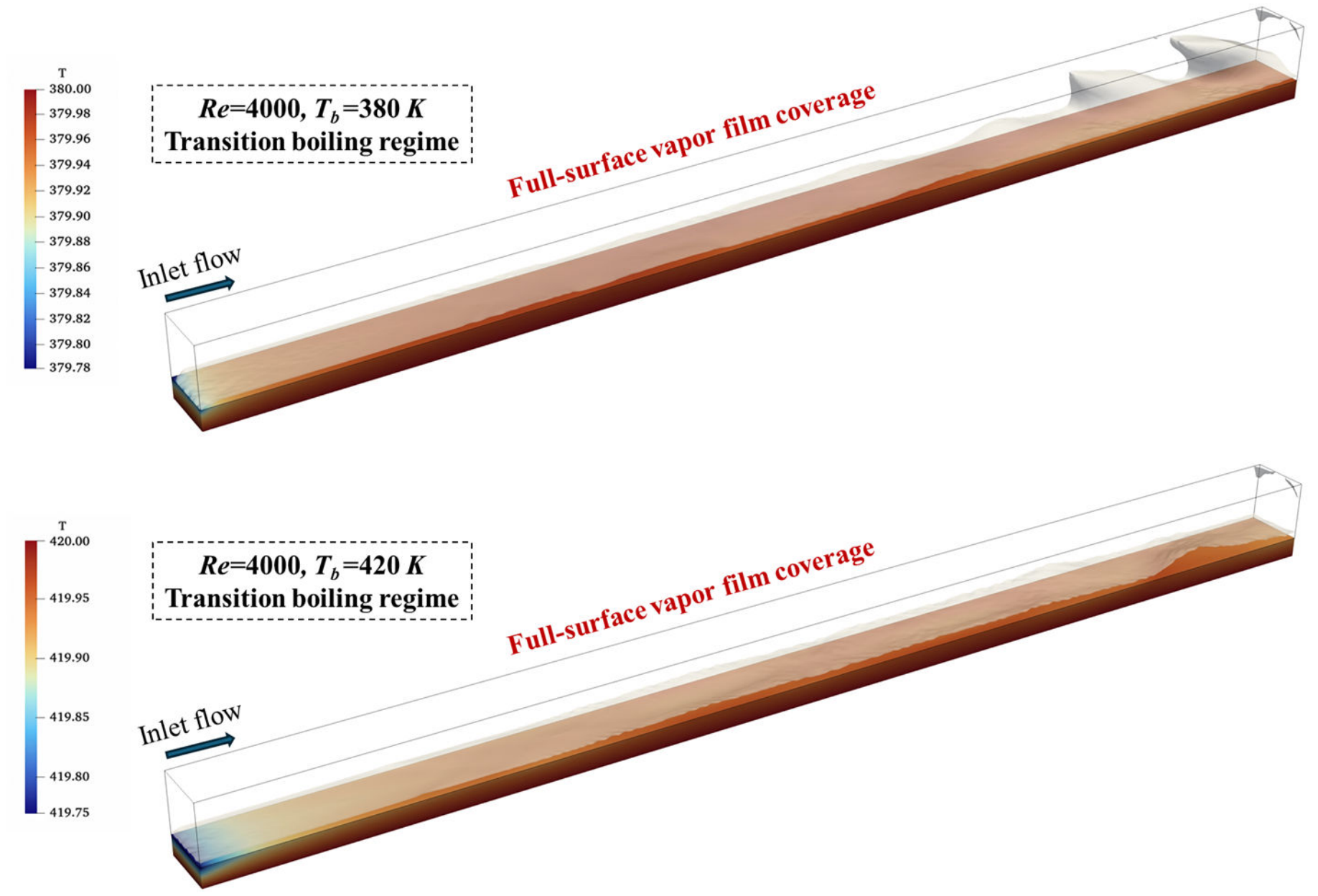

Fig.14 High temperature evolution of the convection dominated elongated vapor mode for *Re*=4000: development of full surface vapor film coverage

At sufficiently high $T_b$, the localized vapor films continue to expand and eventually merge into a nearly continuous vapor film covering almost the entire heated surface. As shown in Fig.14, by $T_b$=380 K the wall is already almost fully covered by a vapor film, and this full surface vapor film coverage persists at even higher wall temperatures such as $T_b$=420 K. This behavior demonstrates that the final stage of the convection dominated elongated vapor mode is a film boiling like state established through the streamwise growth and connection of localized vapor films, rather than through the gradual axial expansion of downstream dryout.

The effect of *Re* within this mode family is further illustrated by comparing the cases with *Re*=3000, 4000, and 5000 at the same wall temperature, as shown in Fig.15 and Fig.16. At $T_b$=355 K, increasing the *Re* clearly intensifies the streamwise elongation of vapor slugs and promotes their earlier conversion into localized vapor films. For *Re*=3000, the dominant structures are still elongated vapor slugs, whereas for *Re*=4000 and especially *Re*=5000, localized vapor films become increasingly evident.

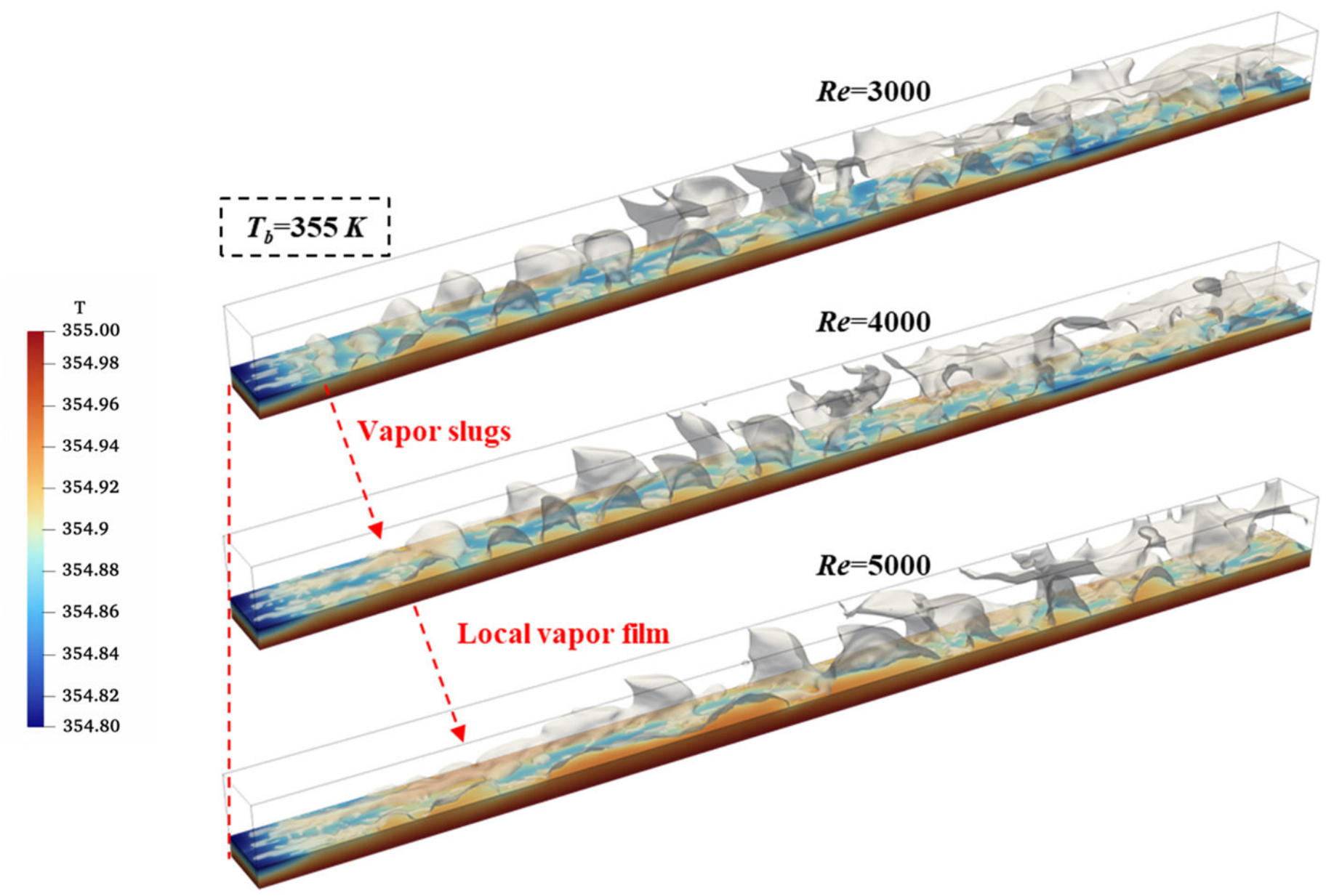

Fig.15 Effect of *Re* on vapor slug elongation and local vapor film formation for $T_b$=355 K

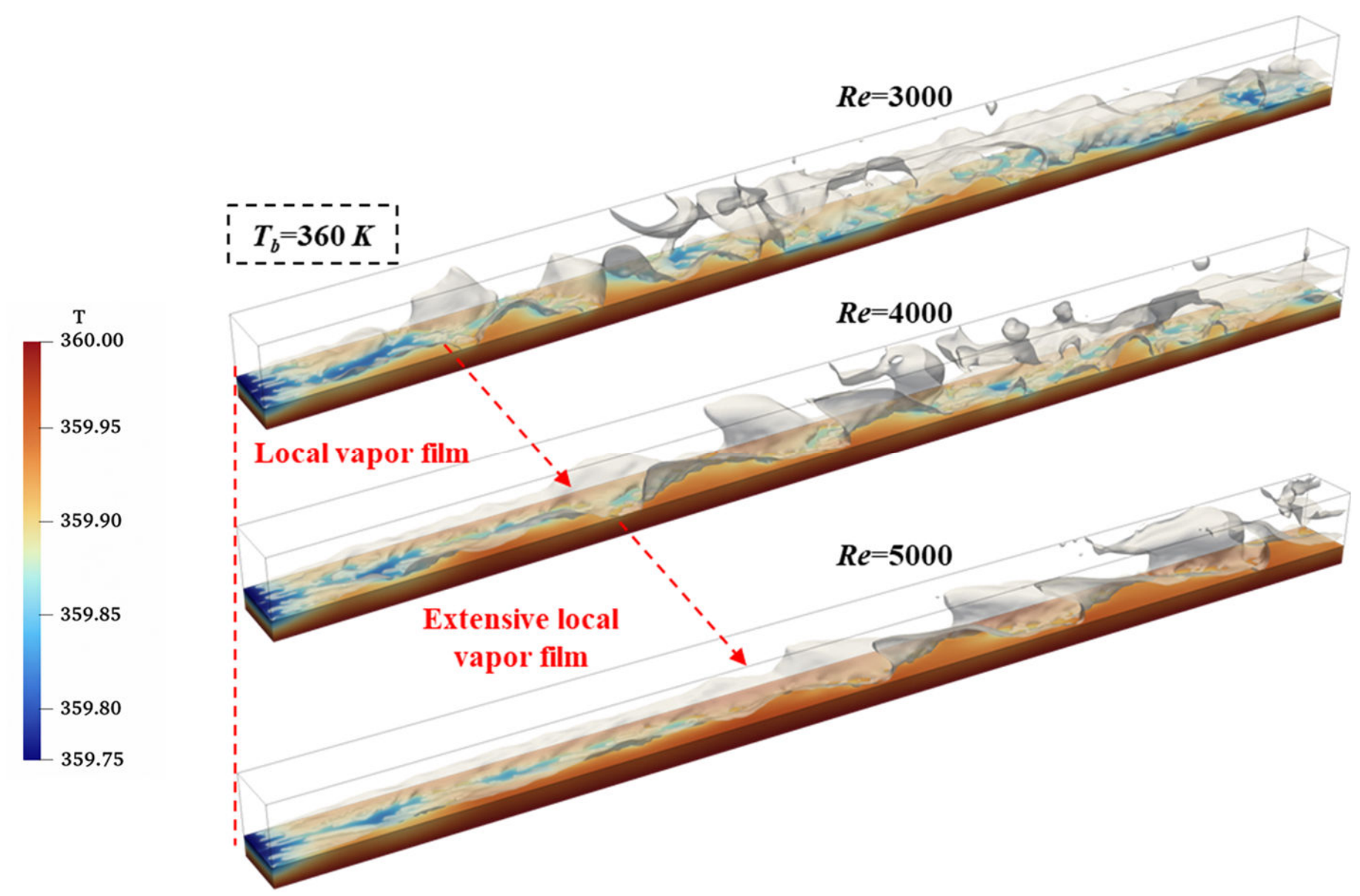

Fig.16 Effect of *Re* on the extension of localized vapor films for $T_b$=360 K

The same trend becomes even more pronounced at $T_b$=360 K, where the increase in $Re$ leads to a substantial extension of the local vapor film and drives the system closer to a local film boiling state. These comparisons confirm that, within the present transition mode, a higher $Re$ accelerates the evolution from elongated vapor slugs to localized vapor films and further promotes the streamwise expansion of film coverage.

The above transition process is directly reflected in the relationship between the time-averaged heat flux $\bar{Q}$ and the wall dryout fraction at the conjugate superheated surface $f_{dry}$. As shown in Fig.17, the $\overline{\mathrm{CHF}}$ decreases from approximately 9.63 kW/m$^2$ at $Re$=3000 to 8.63 kW/m$^2$ at $Re$=4000 and 8.02 kW/m$^2$ at $Re$=5000. Meanwhile, the dryout fraction increases more rapidly and approaches unity at lower $T_b$ as the $Re$ increases. This trend indicates that, although stronger inlet flow enhances liquid replenishment, it simultaneously intensifies convective vapor stretching and promotes the earlier formation and extension of localized vapor films. Consequently, the wall loses effective wetting at lower $T_b$, leading to an earlier deterioration of heat transfer and a lower $\overline{\mathrm{CHF}}$. In other words, for the present high $Re$ cases, the deterioration of boiling heat transfer remains strongly correlated with the increase in dryout fraction, but the physical origin of dryout is different from that in the low $Re$ regime: here it is induced primarily by the convective reorganization of elongated vapor structures into wall-attached vapor films.

It should be noted that the dryout fraction at the time-averaged CHF is not a fixed threshold. In the high $Re$ regime, stronger inlet flow promotes the streamwise elongation of vapor structures and the earlier formation of localized wall-attached vapor films. These connected vapor-film regions can substantially weaken local wall wetting and suppress nucleate boiling heat transfer even when the global dryout fraction is still relatively small. Therefore, with increasing $Re$, the heat flux peak may occur at a smaller dryout fraction, indicating that the morphology and spatial connectivity of near-wall vapor structures are also important for heat transfer deterioration.

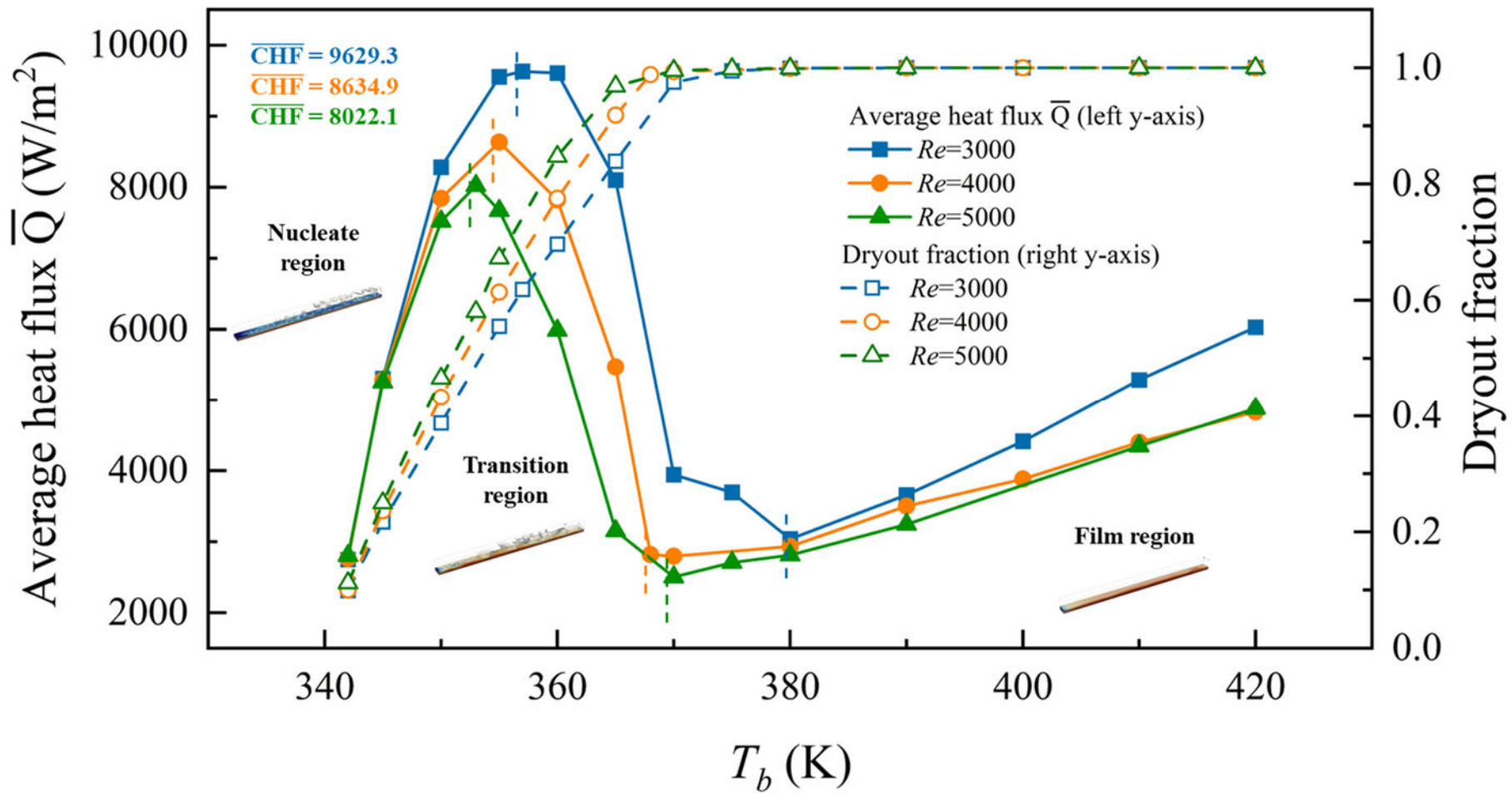

Fig.17 Time-averaged heat flux and wall dryout fraction at conjugate superheated surface for the convection dominated elongated vapor transition mode for *Re*=3000~5000

At higher $T_b$, the heat flux exhibits a slight recovery for all three *Re*, whereas the dryout fraction remains close to unity. Similar to the low *Re* regime, this recovery should not be interpreted as an actual restoration of efficient boiling heat transfer, but rather as a consequence of the increased imposed temperature difference under nearly full vapor film coverage. Therefore, the cases with *Re*=3000~5000 can be classified into the same convection dominated elongated vapor transition family. The increase in *Re* mainly accelerates vapor slug elongation, promotes earlier local film formation, and advances the transition toward full surface film boiling.

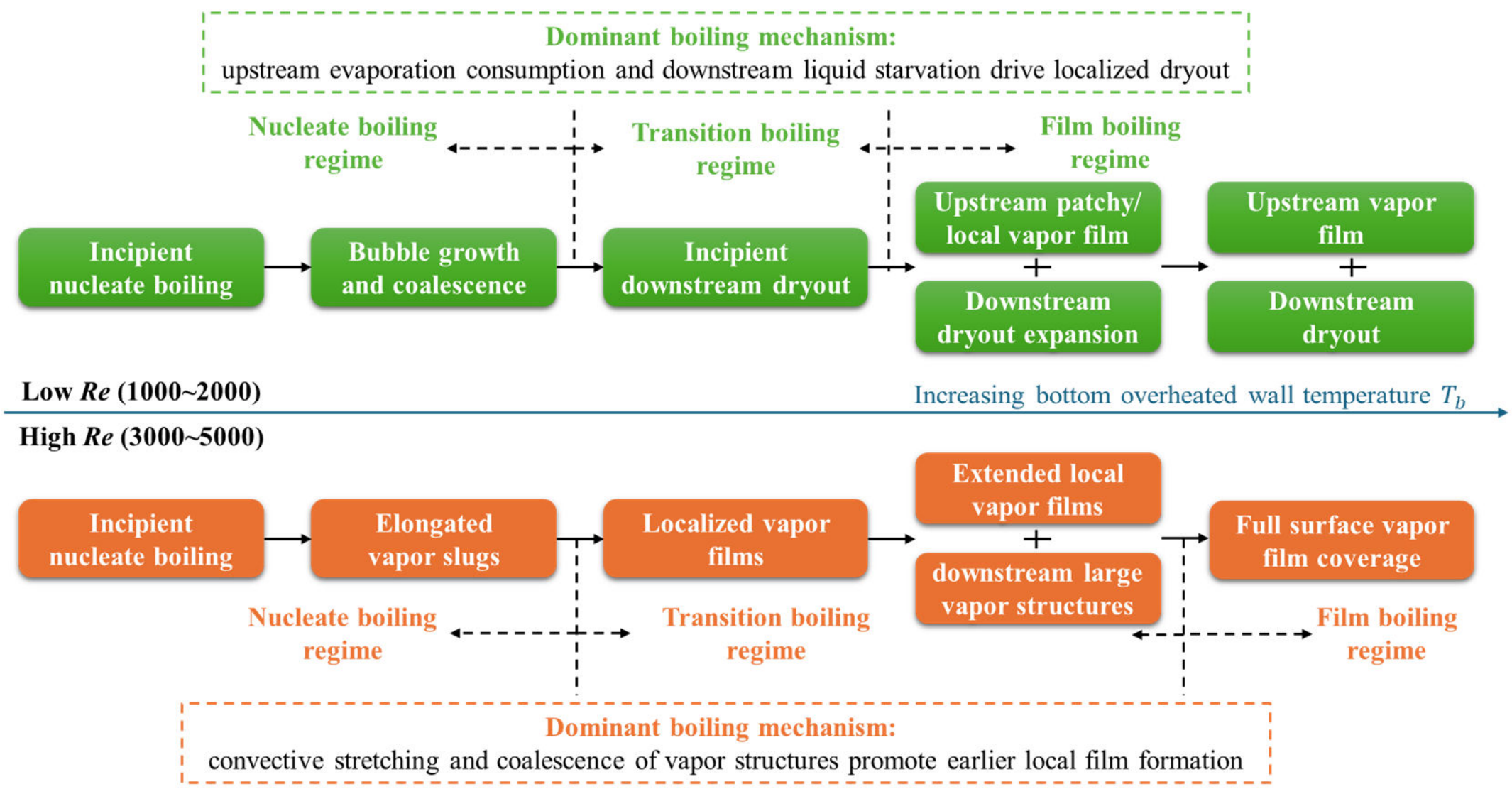

Fig.18 Schematic summary of *Re*-dependent boiling transition pathways

The above results indicate that the boiling transition in the present rectangular mini-channel can be classified into two distinct *Re*-dependent pathways, as summarized schematically in Fig.18. At low *Re* (*Re*=1000~2000), the transition is mainly associated with localized dryout development. In this pathway, boiling initially proceeds through nucleate boiling with bubble growth and coalescence, followed by incipient downstream dryout, and then evolves into a transition boiling state characterized by the coexistence of upstream patchy/local vapor films and downstream dryout expansion. At sufficiently high $T_b$, the channel finally exhibits a high temperature configuration consisting of an upstream vapor film region and a downstream dryout region. In contrast, at higher *Re* (*Re*=3000~5000), the transition is mainly characterized by convective vapor stretching and reorganization. In this case, discrete vapor structures are rapidly elongated into vapor slugs, which subsequently merge into localized vapor films. With further increase in $T_b$, these local vapor films expand continuously along the streamwise direction and eventually develop into nearly full surface vapor film coverage. Therefore, within the present configuration, the two pathways show different dominant features: the low *Re* cases are associated with upstream evaporation consumption and downstream liquid starvation, whereas the high *Re* cases are associated with convective elongation, coalescence, and filmwise reorganization of vapor structures.

**B. Global heat transfer characteristics**

Having identified the two *Re*-dependent transition pathways in Section A, the associated global heat transfer response is examined here using the time-averaged heat flux $\bar{Q}$, the time-averaged Nusselt number $Nu_{ave}$, the wall dryout fraction $f_{dry}$, and the volumetric vapor fraction $\alpha_v$. Both $\bar{Q}$ and $Nu_{ave}$ are obtained by first performing a spatial average over the conjugate boiling surface and then taking the temporal average over the statistically stable stage.

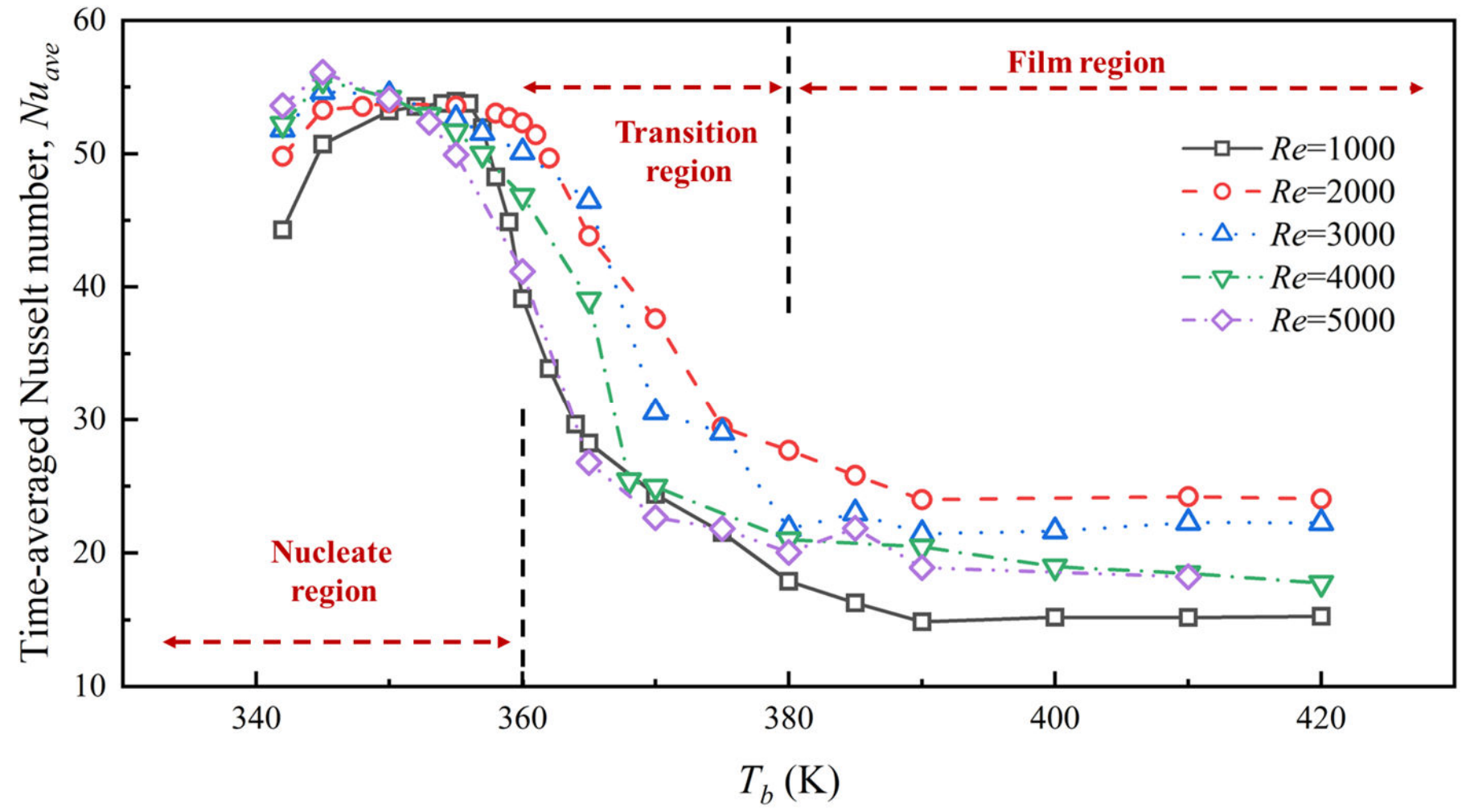

Fig.19 Variation of the time-averaged Nusselt number at the conjugate superheated surface with bottom wall temperature for different inlet *Re*

Fig.19 shows the variation of the time-averaged Nusselt number, $Nu_{ave}$, with bottom wall temperature, $T_b$, for the two different boiling transition pathways. Here, $Nu_{ave}$ characterizes the convective heat transfer capability at the conjugate superheated surface and is closely related to the near-wall vapor distribution and wall wetting condition. In the nucleate boiling regime, increasing $Re$ generally enhances the overall convective heat transfer, as reflected by the larger $Nu_{ave}$, because stronger inlet flow improves both convective transport and liquid replenishment before intense boiling develops. In the fully developed nucleate boiling regime, however, the $Nu_{ave}$ values for different $Re$ collapse into a relatively narrow range of about 52-56, indicating that heat transfer is then mainly associated with vigorous phase change rather than by the inlet flow difference alone. Once the system enters transition boiling, the trend becomes strongly pathway-dependent. For the localized dryout dominated pathway, increasing $Re$ significantly suppresses the extent of dryout, reduces the near wall vapor occupation, and therefore maintains a higher average convective heat transfer coefficient. As a result, $Nu_{ave}$ increases markedly with $Re$ within this regime. By contrast, for the high $Re$ pathway characterized by convective vapor stretching and reorganization, a further increase in $Re$ promotes the elongation of vapor structures and accelerates the formation of localized vapor films, thereby intensifying wall dewetting and causing the system to enter a film boiling dominated state earlier. Consequently, the dryout severity becomes higher and $Nu_{ave}$ decreases with increasing $Re$ in this regime.

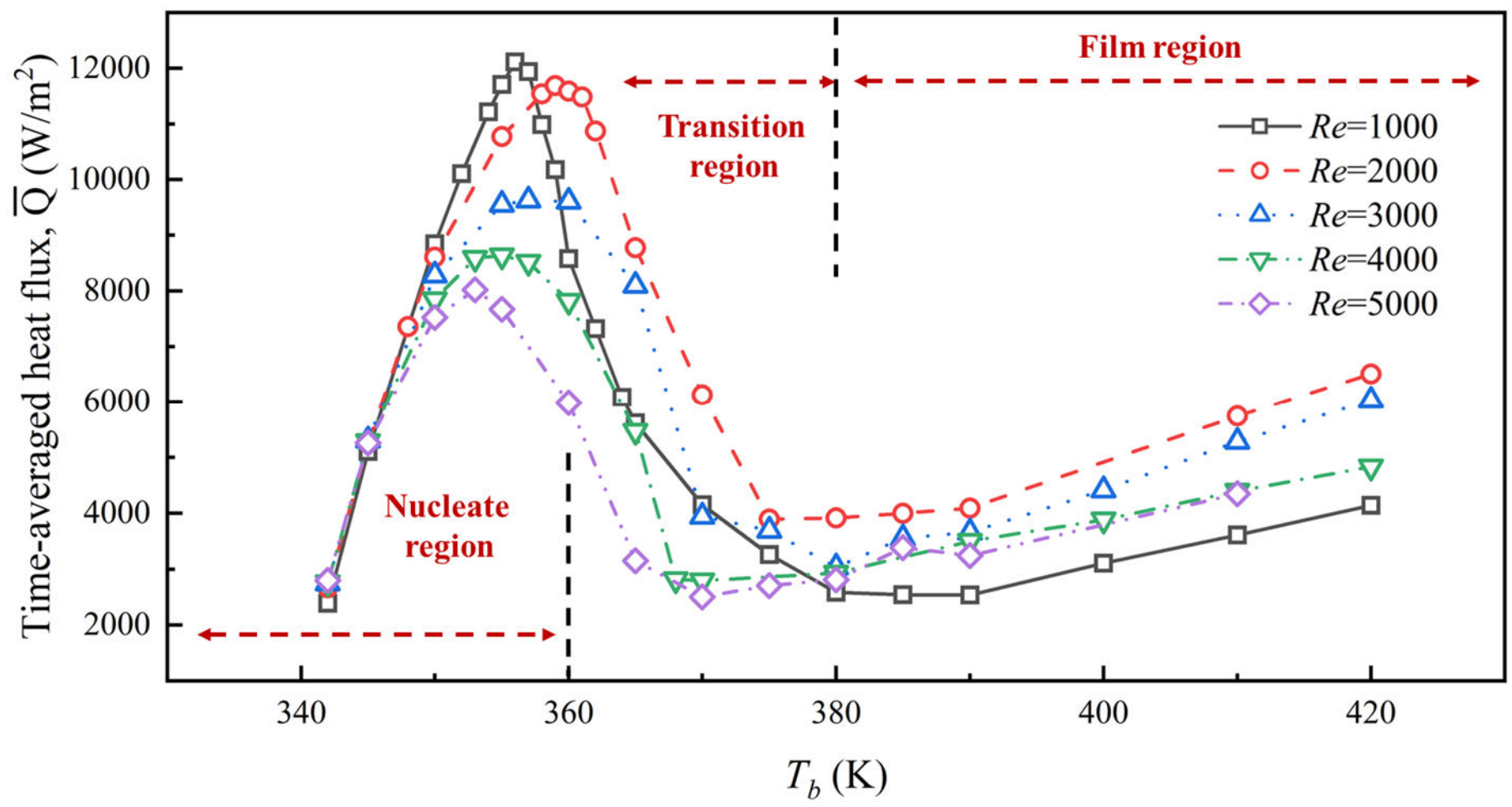

Fig.20 Time-averaged boiling curves for different inlet $Re$

Similarly, the time-averaged wall heat flux $\bar{Q}$ exhibits a comparable trend and is also strongly influenced by the two boiling transition pathways. Fig.20 presents the boiling curves for different $Re$. The low $Re$ cases, which follow the localized dryout dominated pathway, reach higher peak heat fluxes, with time averaged critical heat flux $\overline{\mathrm{CHF}}$ values of approximately 12.1 and 11.7 kW/m$^2$ for $Re$=1000 and 2000, respectively. This is because, under this pathway, $\overline{\text{CHF}}$ is reached while the channel is still in a fully developed nucleate boiling state, as illustrated in Fig.6 and 10, without the presence of pronounced local vapor films or extensive dry patches. Since the present study considers saturated boiling, the contribution of inlet convection remains relatively limited at lower wall superheats, and the wall heat flux is governed mainly by the phase change intensity. Under such conditions, a lower *Re* allows the liquid to reside longer in the channel and undergo more complete vaporization, which explains why the $\overline{\text{CHF}}$ for *Re*=1000 is slightly higher than that for *Re*=2000. In contrast, the high *Re* cases, which follow the convection dominated elongated vapor pathway, exhibit markedly lower $\overline{\text{CHF}}$ values, decreasing from about 9.6 to 8.0 kW/m$^2$ as *Re* increases, which are clearly lower than those of the low *Re* cases.

This reduction in $\overline{\text{CHF}}$ with increasing *Re* in the high *Re* family is physically important. Although a larger *Re* strengthens inlet liquid replenishment, it also intensifies convective stretching of vapor structures, as shown in Figs.12, 15, and 16. Higher *Re* promotes the transformation of elongated vapor slugs into localized vapor films and accelerates the streamwise extension of these films. Consequently, effective wall wetting is lost at lower wall temperatures, so the wall enters a deteriorated boiling state earlier and the $\overline{\text{CHF}}$ decreases instead of increasing. In other words, once convective vapor elongated becomes dominant, further increasing *Re* no longer benefits $\overline{\text{CHF}}$, because the earlier formation and expansion of wall-attached vapor films outweigh the positive effect of enhanced liquid supply. Once the system enters the transition boiling and film boiling regimes, the evolution of the boiling curves becomes broadly consistent with that of $Nu_{ave}$. A larger *Re* intensifies the extension of local vapor films and therefore results in poorer heat transfer performance.

This $\overline{\text{CHF}}$ behavior further reflects the competition between these mechanisms. Among all cases, *Re*=2000 provides the most favorable overall thermal response in the high temperature range: although its peak heat flux is slightly lower than that of *Re*=1000, its $Nu_{ave}$ remains noticeably higher after the onset of transition boiling, and its heat flux recovery in the film boiling regime is also the strongest. This suggests that *Re*=2000 offers the best compromise between maintaining liquid replenishment and delaying large scale heat transfer deterioration. By comparison, *Re*=1000 suffers more severe local dryout after $\overline{\text{CHF}}$, whereas *Re*=4000 and 5000 deteriorate earlier because elongated vapor structures are more readily converted into localized vapor films and then into nearly full surface vapor coverage.

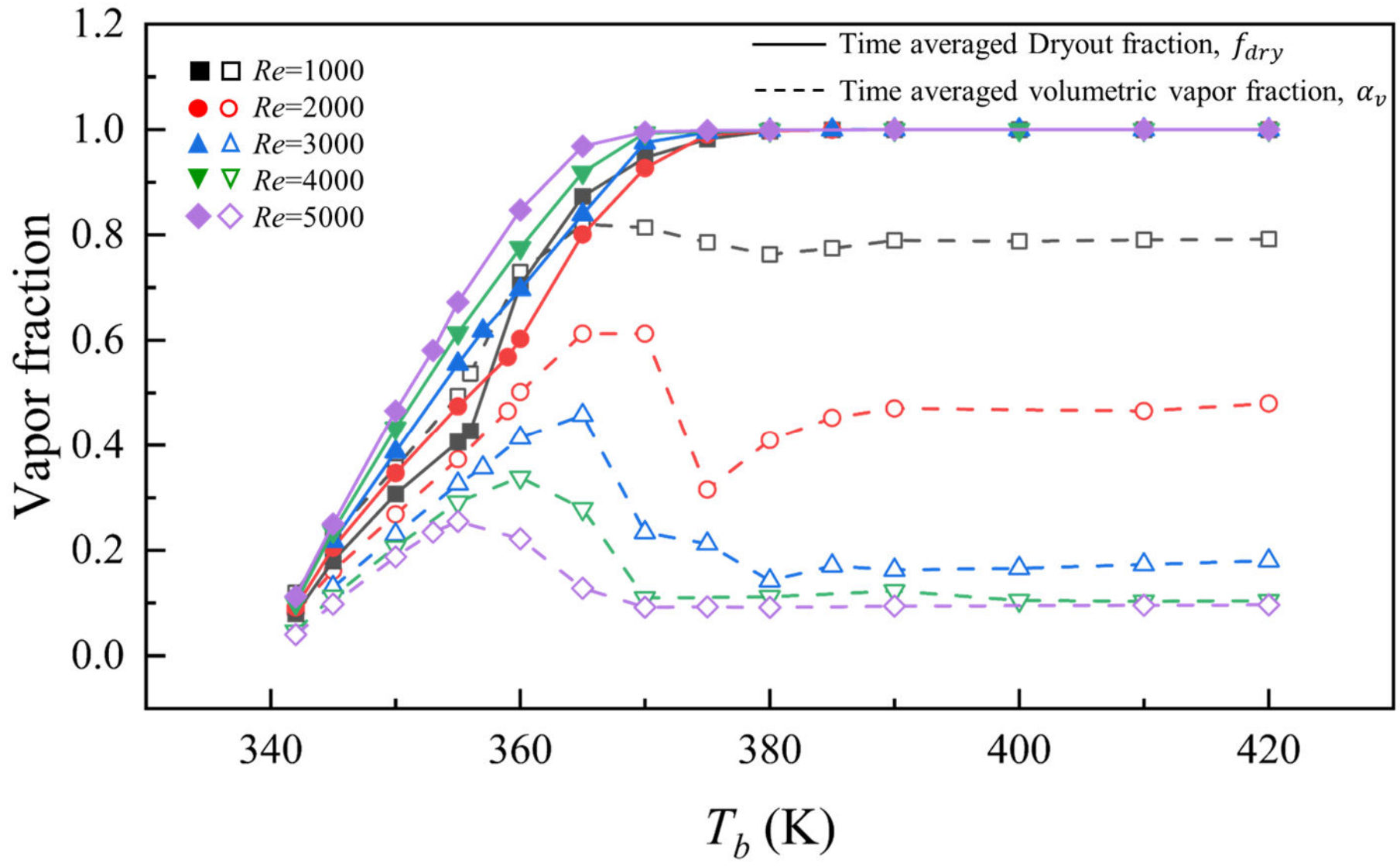

Fig.21 Variation of the time-averaged conjugate superheated surface dryout fraction and volumetric vapor fraction with bottom wall temperature for different inlet *Re*

The relationship between heat transfer deterioration and the two phase flow characteristics is further illustrated in Fig.21. For all cases, the wall dryout fraction $f_{dry}$ increases monotonically with $T_b$ and eventually approaches unity, indicating that the long term degradation of boiling heat transfer is associated with the loss of effective wall wetting. By contrast, the volumetric vapor fraction $\alpha_v$ shows very different trends for the two transition families. In the low *Re* cases, especially at *Re*=1000, $\alpha_v$ remains relatively high even when $f_{dry}$ is already close to 1, indicating strong vapor retention inside the channel. In the high *Re* cases, however, $f_{dry}$ reaches unity at relatively low wall temperatures, while $\alpha_v$ decreases to much lower values in the high $T_b$ range, because the vapor is continuously stretched and convected downstream rather than accumulating in the channel. Therefore, $f_{dry}$ is more directly correlated with heat transfer deterioration than $\alpha_v$, whereas $\alpha_v$ mainly reflects the extent of vapor holdup in the channel.

Overall, Fig.19-21 show that the global heat transfer behavior is jointly determined by the boiling transition pathway and the associated wall wetting condition. The localized dryout dominated mode at low *Re* allows a higher $\overline{\mathrm{CHF}}$ to be reached, but may suffer severe post peak deterioration once dryout expands. The convection dominated elongated vapor mode at high *Re* promotes earlier local film formation and extension, thereby reducing $\overline{\mathrm{CHF}}$ and accelerating the transition toward film boiling dominated conditions. These results suggest that, in the present rectangular mini-channel configuration, inlet *Re* is not only a hydrodynamic parameter but also an important factor affecting the coupling among vapor structure evolution, wall dryout, and global heat transfer performance.

It should be emphasized that the above discussion is based on globally time-averaged heat transfer metrics. Although the $\overline{\mathrm{CHF}}$ decreases with increasing *Re*, the localized dryout dominated mode at low *Re* is accompanied by extensive dryout over large portions of the heated surface, which implies a substantial thermal risk in practical applications.

## V. CONCLUSION

In this study, 3D conjugate simulations were conducted to investigate saturated flow boiling in a rectangular mini-channel. The effects of inlet *Re* on boiling transition and heat transfer performance were systematically examined using a C++ based open-source finite-volume framework OpenFOAM-v2506, in which a spatial nucleation model was incorporated to reproduce realistic boiling processes. The main conclusions are as follows:

1. **Two distinct Reynolds number dependent boiling transition pathways were identified.**

   At low *Re* (*Re*=1000~2000), the boiling transition is mainly associated with the initiation, growth, and lateral expansion of localized dryout patches. At higher *Re* (*Re*=3000~5000), the transition is mainly characterized by convective stretching and downstream reorganization of vapor structures, which promotes the formation and extension of localized wall-attached vapor films.

2. **These two transition pathways are associated with different heat transfer deterioration behaviors.**

   For the low *Re* pathway, $\overline{\mathrm{CHF}}$ is reached while the channel is still in a fully developed nucleate boiling state, so the corresponding time-averaged critical heat flux remains relatively high. For the high *Re* pathway, stronger vapor elongation and earlier local film formation cause the wall to enter a deteriorated boiling state, thereby reducing the heat transfer capability in the transition and film boiling regimes.

3. **The global heat transfer performance shows a clear Reynolds number dependence.**

   The low *Re* cases reach higher time-averaged critical heat fluxes, with $\overline{\mathrm{CHF}}$ values of about 12.1 and 11.7 kW/m$^2$ for *Re*=1000 and 2000, respectively. In contrast, for the higher $Re$ family, $\overline{\mathrm{CHF}}$ decreases from about 9.6 kW/m$^2$ at $Re$=3000 to 8.0 kW/m$^2$ at $Re$=5000. The variation of the time-averaged Nusselt number $Nu_{ave}$ exhibits the similar pathway dependence. Among the cases considered, $Re$=2000 provides the best overall compromise in the high wall temperature range: although its $\overline{\mathrm{CHF}}$ is slightly lower than that of $Re$=1000, it maintains a higher $Nu_{ave}$ after the onset of transition boiling and shows the strongest heat flux recovery in the film boiling regime.

Overall, the present results indicate that, for the rectangular mini-channel configuration, working fluid, boundary conditions, and operating range considered in this study, $Re$ acts as an important mode selection parameter that strongly affects vapor structure evolution, boiling transition, wall dryout, and heat transfer performance. It should be emphasized that other nondimensional groups associated with wall superheat, surface tension, inertia, viscosity, confinement, and fluid properties may also influence boiling transition when the geometry, fluid, surface condition, or heating condition is changed. Therefore, the $Re$-dependent transition pathways identified here should be interpreted within the scope of the present configuration and $T_b$ range, rather than as a universal regime criterion for all mini-/micro-channel boiling systems. Nevertheless, the present study still has several limitations. In particular, subcooled boiling has not been considered, and phase change on the realistic vertical sidewalls (fins/ribs) has not yet been included. Despite these limitations, the present work establishes a physically consistent baseline for subsequent studies on boiling control in mini- and micro-channels using global or localized electric fields.

**ACKNOWLEDGMENTS**

1. Author Q. Wang supported by the Heilongjiang Province Postdoctoral General Funding (Grant No. AUGA4110006724) and Postdoctoral research start-up funds, HIT (Grant No. AUGA5710027524).
2. Author Alberto T. Pérez acknowledges financial support from the Spanish Ministerio de Ciencia e Innovación under grant number PID2021-123297NB-I00.

# 参考文献